\documentclass{article}

\usepackage{PRIMEarxiv}

\usepackage[utf8]{inputenc} 
\usepackage[T1]{fontenc}    
\usepackage{hyperref}       
\usepackage{url}            
\usepackage{booktabs}       
\usepackage{amsfonts}       
\usepackage{nicefrac}       
\usepackage{microtype}      
\usepackage{lipsum}
\usepackage{fancyhdr}       
\usepackage{graphicx}       
\graphicspath{{media/}}     
\usepackage{graphicx}
\usepackage{subcaption}
\usepackage{multirow}
\usepackage{enumitem}
\usepackage{caption}
\usepackage{float}
\usepackage{amsmath}

\title{Beamforming Feedback as a Novel Attack Surface for Wi-Fi Physical-Layer Security
}

\author{
 Jingzhe Zhang \\
  University of South Florida\\
  \texttt{jingzhe@usf.edu} \\
   \And
 Yitong Shen \\
  University of South Florida\\
  \texttt{shen202@usf.edu} \\
  \And
 Ning Wang \\
  University of South Florida\\
  \texttt{ningw@usf.edu} \\
\And
 Yili Ren \\
  University of South Florida\\
  \texttt{yiliren@usf.edu} \\
}

\begin{document}
\maketitle

\begin{abstract}
With the rapid evolution of wireless technologies, Wi-Fi has expanded beyond its original role in data transmission to support various emerging applications, particularly in physical-layer security, including device authentication, user authentication, and secret key generation. Despite extensive research on Wi-Fi Channel State Information (CSI)-based physical-layer security, its vulnerabilities remain largely unexplored. In this work, we propose BFIAttack, a novel attack that exploits Beamforming Feedback Information (BFI) to reconstruct the CSI of a legitimate user or device, thereby compromising Wi-Fi-based physical-layer security. We realize the attack by leveraging a closed-form CSI reconstruction method for the single-antenna station scenario and a maximum likelihood estimation-based CSI reconstruction for the multi-antenna station scenario. Moreover, we exploit spatial similarities among antenna pairs to refine the reconstructed CSI and enhance attack effectiveness. Experimental results show that BFIAttack achieves an average attack success rate of $73\%$ in multi-antenna station scenarios with no more than five attack attempts, and over $93\%$ in single-antenna station scenarios with only a single attempt. BFIAttack reveals critical vulnerabilities in existing Wi-Fi-based physical-layer security.
\end{abstract}


\section{Introduction}

As wireless technology continues to advance, Wi-Fi, one of the most widely deployed communication technologies, has evolved beyond traditional data transmission to support a broad range of emerging applications~\cite{ma2019wifi}, particularly in physical-layer security. Wi-Fi-based physical-layer security extracts distinctive features of devices and users from Channel State Information (CSI)~\cite{halperin2011tool}, which captures the unique effects of signal reflection, scattering, and attenuation~\cite{tan2022commodity}. These features can enable essential security applications such as device authentication~\cite{chen2021enhancing, kong2024csi, senigagliesi2020comparison, baracca2012physical, jiang2013rejecting, liu2017authenticating}, user authentication~\cite{zeng2016wiwho, shi2021wifi, zhang2016wifi, zhang2020gate, zhang2022metaganfi}, and secret key generation~\cite{liu2013fast, li2022fast, du2024secure, ribouh2020channel}.
They offer a non-cryptographic, low-cost solution that is particularly well-suited for power- and resource-constrained Internet of Things (IoT) and wireless devices, providing a promising pathway to enhance the security of IoT, mobile, and wireless systems.
Moreover, Wi-Fi's ubiquity and the ability to reuse existing wireless infrastructure offer a widely deployable and cost-effective foundation for implementing physical-layer security applications across diverse environments, including homes, offices, and public spaces. Consequently, Wi-Fi-based physical-layer security has attracted significant research attention in recent years~\cite{chen2021enhancing, kong2024csi, baracca2012physical, liu2013fast, du2024secure, wang2022caution, kong2023toward, ren2023person, xin2016freesense, shi2017smart, chen2025scattershield}.  


However, the vulnerabilities of Wi-Fi-based physical-layer security remain largely unexplored. Some existing approaches employ random attacks~\cite{shi2021wifi}, where adversaries attempt to spoof physical-layer security applications (e.g., authentication) by generating random Wi-Fi signals. Such attacks are generally weak, as the probability of randomly generating Wi-Fi signals that closely resemble those of legitimate users or devices is extremely low.
Other approaches investigate knowledgeable attacks~\cite{fang2016virtual, fang2014you}, which assume that adversaries have direct access to the CSI of legitimate users or devices. Although such an attack may be effective in controlled scenarios, it is impractical in real-world settings, as it typically necessitates compromising or controlling the user's device.
DomPathCon~\cite{qu2024guessing} is a potential attack against Wi-Fi-based physical-layer security, in which the adversary actively attempts to guess the CSI of a legitimate user or device. However, it suffers from limited stealth, as it requires a large number of attack attempts (typically more than 50 attempts) to achieve a reasonable attack success rate, which significantly increases the likelihood of detection by the user. The number of required attempts further grows with both signal bandwidth and the number of antennas, making the attack less feasible for modern Wi-Fi that employs wider bandwidths and multiple antennas. 


Wi-Fi Beamforming Feedback Information (BFI)~\cite{haque2023wi, yi2024bfmsense}, a partial and compressed form of CSI originally designed to facilitate beamforming and enhance communication performance, presents a new opportunity to launch novel attacks on Wi-Fi-based physical-layer security. In particular, BFI is standardized in IEEE 802.11ac/ax~\cite{6687187, 9442429}, and is supported by nearly all modern Wi-Fi devices~\cite{wu2023enabling}.
However, BFI is transmitted in plaintext and can be easily captured by adversaries through passive packet sniffing using commodity Wi-Fi devices without requiring any firmware or hardware modifications~\cite{li2024efficient}. These characteristics pose significant risks to Wi-Fi-based physical-layer security and threaten a broad range of IoT, wireless, and mobile systems. Therefore, investigating the adversarial potential of BFI and uncovering these overlooked vulnerabilities is both urgent and essential.


In this work, we present BFIAttack, a novel BFI-based attack that reveals critical vulnerabilities in Wi-Fi-based physical-layer security. BFIAttack is stealthy, as it can compromise multiple security applications with only a few attack attempts (e.g., no more than 5 attempts), significantly reducing the likelihood of user detection. It is also practical, as it requires only passive Wi-Fi packet sniffing, without requiring direct access to a legitimate user's CSI or control over the user's devices. Furthermore, BFIAttack is device-agnostic, functioning effectively across various devices with different bandwidths and numbers of antennas.
The core idea of BFIAttack is to leverage BFI, passively captured over the air via a sniffer, to reconstruct the legitimate user's or device's CSI. This enables the adversary to compromise various Wi-Fi-based physical-layer security applications, such as impersonating the legitimate user/device or recovering a secret key.


However, realizing BFIAttack is non-trivial, as it requires overcoming several key challenges. 
First, although BFI is derived from CSI, it undergoes an irreversible transformation via Singular Value Decomposition (SVD)~\cite{wu2023enabling}, rendering the direct recovery of CSI from BFI infeasible.
To address this challenge, we divide the problem into two scenarios and tackle each separately: one where the Wi-Fi station (STA) is equipped with a single antenna, and another where the STA is equipped with multiple antennas.
We note that the Wi-Fi access point (AP) can have an arbitrary number of antennas. 
In the single-antenna STA scenario, we exploit a unique mathematical relationship between CSI and BFI to derive a closed-form CSI reconstruction, enabling highly effective attacks.
In contrast, no such closed-form solution exists in the multi-antenna STA scenario. To overcome this, we leverage a computationally efficient maximum likelihood estimation (MLE) method to infer channel parameters and thus reconstruct the CSI required for the attack.

Second, in the multi-antenna STA scenario, we observe that the MLE problem is inherently non-convex, giving rise to multiple maxima. As a result, multiple valid MLE solutions may exist, each corresponding to a reconstructed CSI. To identify those reconstructed CSIs most likely corresponding to the legitimate user/device, we introduce theory-practice dual constraints, which eliminate CSIs that are inconsistent with IEEE 802.11ac/ax standards and real-world physical conditions. In particular, we derive the theoretical amplitude range of the reconstructed CSI based on the formulation of BFI as defined in the IEEE 802.11ac/ax standards. Furthermore, the reconstructed CSI must exhibit channel parameters that are consistent with the physical characteristics of real-world signal propagation. 




Third, effective refinement of the reconstructed CSI is essential for ensuring successful attacks, especially in the multi-antenna STA scenario where no closed-form solution exists. We propose a CSI refinement method that leverages spatial similarities among STA-AP antenna pairs to enhance attack efficacy.
Specifically, antennas on IoT, mobile, and wireless devices (e.g., laptops and smart home hubs) are typically placed in close proximity (e.g., a few centimeters apart), and the transmitted signals propagate through nearly identical environments. These conditions induce strong correlations in signal attenuation across different STA-AP antenna pairs~\cite{mei2020envelope, nosrat2011mimo}. Our refinement method exploits these correlations to further fine-tune the reconstructed CSI, thereby increasing the likelihood of a successful attack.

We evaluate BFIAttack against multiple Wi-Fi-based physical-layer security applications, including device authentication, user authentication, and secret key generation. Experiments are conducted across diverse real-world environments such as laboratories, apartments, and outdoor settings. To further assess BFIAttack, we conduct experiments under more challenging settings, including varying distances between the adversary's sniffer and the user's devices, different commodity Wi-Fi devices, different bandwidths and numbers of antennas, non-line-of-sight (NLoS) conditions, etc. 
Experimental results demonstrate that BFIAttack achieves an average attack success rate of $73\%$ in multi-antenna STA scenarios using no more than five attack attempts, and over $93\%$ in single-antenna STA scenarios with only one attack attempt. 
The main contributions of our work are summarized as follows:

\begin{itemize}
    \item We propose BFIAttack, a novel attack against Wi-Fi-based physical-layer security. 
    By leveraging passively captured BFI, BFIAttack can reconstruct the legitimate user's or device's CSI, thereby compromising various security applications. This reveals new vulnerabilities of Wi-Fi-based physical-layer security.
    \item We leverage the underlying relationship between BFI and CSI to develop a closed-form CSI reconstruction method for attacking single-antenna STA scenarios. For multi-antenna STA scenarios, we utilize an MLE-based CSI reconstruction approach and further design a refinement method that exploits spatial similarities among antenna pairs to improve attack success rates.
    \item We conduct extensive evaluations for our attack across multiple Wi-Fi-based physical-layer security applications and in diverse environmental settings. The results show that the BFIAttack achieves high attack success rates. In addition, we discuss potential countermeasures to mitigate the vulnerabilities exposed by our findings.
\end{itemize}

\section{Related Work}

\textbf{Wi-Fi-Based Physical-Layer Security Applications.} 
In this work, we center on three core applications of Wi-Fi-based physical-layer security, including device authentication, user authentication, and secret key generation.


\textit{Device Authentication.} 
Recently, Wi-Fi CSI-based device authentication primarily relied on machine learning and deep learning algorithms~\cite{pei2014channel, wang2016csideepfi, song2021enhancing, chen2021authenticating, jing2023multi}. For example, Liu et al.~\cite{liu2014practical} extract the CSI to construct device-specific profiles for authentication using the K-means algorithm. Kong et al.~\cite{kong2023physical} employ micro-signals embedded within CSI as fingerprints and apply a K-Nearest Neighbors algorithm for device authentication. Besides, DeepFi~\cite{wang2016csideepfi} uses deep neural network for location distinction of Wi-Fi devices. Chen et al.~\cite{chen2021authenticating} develop a device authentication framework leveraging convolutional neural networks on CSI amplitude. Moreover, Jing et al.~\cite{jing2023multi} adopt a ResNet architecture to extract CSI features and perform authentication based on classification outcomes. Song et al.~\cite{song2021enhancing} further propose an authentication mechanism that combines an autoencoder with the dispersion degree of CSI measurements.

\textit{User Authentication.}
Various features of the human body and activities can be extracted from the Wi-Fi CSI for user (i.e., human) authentication in a device-free manner. Gait and activity-based authentication has attracted considerable research attention~\cite{pokkunuru2018neuralwave, zhang2020gate, wang2021gait, deng2022gaitfi, wang2022caution, lin2018wiau, kong2021multiauth, shi2021wifi, gu2021secure, xin2016freesense}, as human activity patterns provide distinctive motion signatures and can be easily captured by Wi-Fi. In parallel, gesture-based authentication has emerged as another research direction~\cite{kong2019fingerpass, kong2020continuous, li2020wihf, jung2021wi, kong2022push, al2016wiger}. These authentication systems exploit unique fluctuations in CSI induced by human gestures. Some systems have focused on human body-based biometric characteristics (e.g., breathing and body shape)~\cite{xu2017radio, huang2022continuous, wang2019wipin, wu2022widff, lin2023contactless, fang2020eyefi, li2024spacebeat}, which leverage the unique physiological properties of individuals that subtly modulate Wi-Fi signals.

\textit{Secret Key Generation.}
Wi-Fi CSI plays a vital role in secret key generation between devices in the wireless environment. Several studies~\cite{liu2013fast, peng2017secret, ji2021wireless, altun2022scalable, aldaghri2020physical, ribouh2020channel} have utilized CSI for secret key generation. For example, Aldaghri et al.~\cite{aldaghri2020physical} design a secret key generation algorithm that denoises and normalizes CSI and then quantizes it into bit sequences to derive the secret key. Liu et al.~\cite{liu2013fast} propose to combine CSI and channel gain complement to achieve fast secret key generation.






\textbf{Attacks on Wi-Fi-Based Sensing and Security Systems.}
Some prior research has explored attacks on Wi-Fi CSI-based systems~\cite{cao2024security, liu2022physical, liu2023time, jiang2024risiren, xu2022wicam}. However, the majority of these efforts target sensing systems. For example, WiAdv~\cite{zhou2022wiadv} investigates the security vulnerabilities of Wi-Fi-based gesture recognition systems by crafting adversarial signals that induce incorrect predictions. 
Li et al.~\cite{li2024practical} present unnoticeable and universal adversarial attacks on deep learning-enabled Wi-Fi sensing systems by manipulating pilot symbols within packets. RIStealth~\cite{zhou2023ristealth} and RISiren~\cite{jiang2024risiren} demonstrate attacks using reconfigurable intelligent surfaces to render moving people undetectable and to mislead human activity recognition systems, respectively. 
Only a limited number of existing works have explored attacks on Wi-Fi-based physical-layer security systems. For instance, DomPathCon~\cite{qu2024guessing} proposes an attack model that targets CSI-based authentication by guessing the CSI values of dominant propagation paths. However, it requires a large number of attempts to achieve success, particularly for modern Wi-Fi devices equipped with multiple antennas and wide bandwidths, thereby significantly increasing the risk of detection. Thus, the vulnerabilities of Wi-Fi-based physical-layer security remain insufficiently investigated.

\textbf{Beamforming Feedback Information.}
More recently, BFI has gained increasing attention since it can be readily extracted from almost all Wi-Fi devices in a plaintext manner~\cite{wu2023enabling}. Therefore, BFI can facilitate numerous sensing applications~\cite{haque2023wi, li2024efficient, yi2024bfmsense, chen2024echoes, wang2024muki, chen2025beyes, haque2025beamsense, hu2024m, wang2025freebfi, wang2024wi2dmeasure}. For instance, SThief~\cite{chen2024silent} exploits BFI exchanged between point-of-sale (POS) terminals and access points to sense keystrokes on POS keypads. Wi-BFI~\cite{haque2023wi} achieves a human activity recognition system using BFI. M$^2$-Fi~\cite{hu2024m} is a BFI-based respiration monitoring system. BFMSense~\cite{yi2024bfmsense} shows that BFI can be used for fine-grained human motion sensing. BeamSense~\cite{wu2023enabling} leverages bidirectional BFI to estimate multipath channel characteristics for Wi-Fi sensing. FreeBFI~\cite{wang2025freebfi} demonstrates that fine-grained sensing can be achieved using BFI across an arbitrary number of antennas.
However, most existing studies primarily explore the potential of BFI for sensing applications. BeamCraft~\cite{xu2024beamforming} is one of the few works that examine the adversarial potential of BFI, demonstrating how forged beamforming feedback can be used to manipulate Wi-Fi communication traffic. Nevertheless, the broader attack capabilities of BFI, particularly in compromising Wi-Fi-based physical-layer security, remain largely overlooked.

\begin{figure}[t]
  \centering
  \begin{minipage}[t]{0.48\linewidth}
    \centering
    \includegraphics[width=\linewidth]{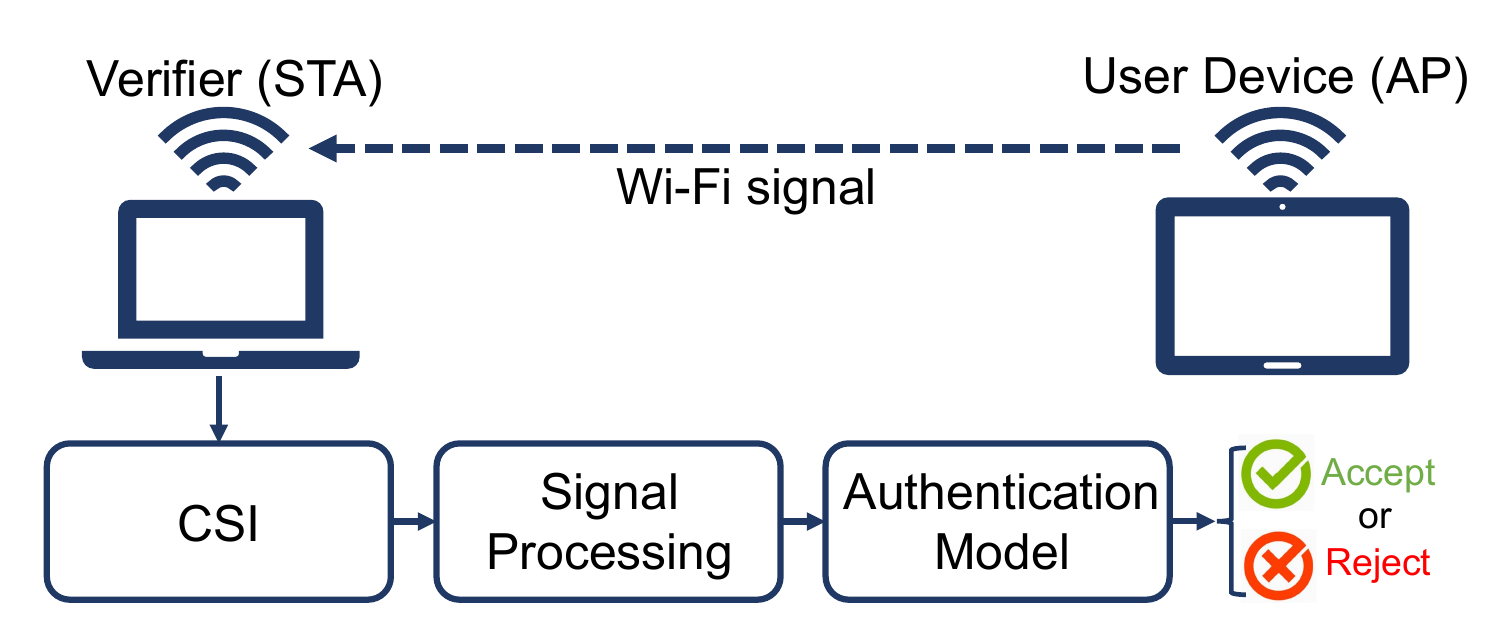}
    \captionof{figure}{Wi-Fi-based device authentication.}
    \label{fig:device-auth}
  \end{minipage}
  \hfill
  \begin{minipage}[t]{0.48\linewidth}
    \centering
    \includegraphics[width=\linewidth]{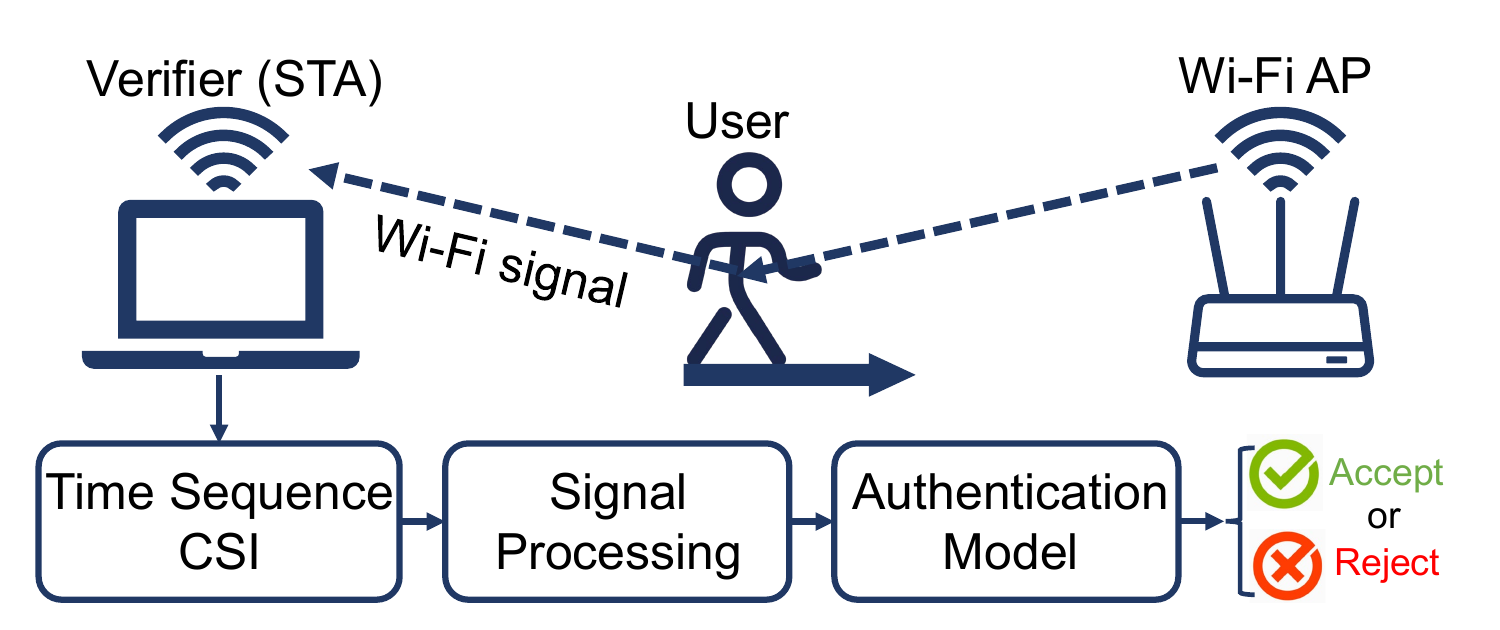}
    \captionof{figure}{Wi-Fi-based user authentication.}
    \label{fig:user-auth}
  \end{minipage}
\end{figure}

\section{Preliminary}

In this section, we first introduce the typical procedures of three representative Wi-Fi-based physical-layer security applications, and then describe the fundamentals of BFI.



\subsection{Typical Procedures of Wi-Fi-Based Physical-Layer Security Applications}

Wi-Fi CSI captures fine-grained channel characteristics between a Wi-Fi AP and a STA at the physical layer. CSI is highly sensitive to environmental conditions, device location, human activities, and multipath propagation~\cite{li2025rethinking, li2025wilife, gao2023wicgesture, hu2025poison, zeng2018fullbreathe, zeng2019farsense}. Thus, it encodes unique signatures associated with devices, users (i.e., humans), and the surrounding environment, thereby enabling a range of security applications such as device authentication, user authentication, and secret key generation.

\textbf{Device Authentication.}
The basic procedure of Wi-Fi-based device authentication is illustrated in Figure~\ref{fig:device-auth}. The verifier (i.e., STA) receives Wi-Fi signals transmitted from the user's device (i.e., AP) and extracts the CSI. A signal processing module then denoises and normalizes the extracted CSI. Subsequently, the verifier utilizes the processed CSI to train an authentication model that characterizes the user device's profile. Such a model can be constructed using machine learning algorithms~\cite{liu2014practical,song2021enhancing,chen2021authenticating,jing2023multi}. Finally, the verifier determines whether the CSI of the incoming signal matches the established device profile.


\textbf{User Authentication.}
A typical Wi-Fi-based user (i.e., human) authentication system is shown in Figure~\ref{fig:user-auth}. The verifier collects a time sequence of Wi-Fi signals transmitted by the AP and then reflected by the human body. Next, it extracts the time sequence CSI from multiple received packets and applies a preprocessing step to remove noises. The verifier trains an authentication model~\cite{ pokkunuru2018neuralwave, kong2019fingerpass, gu2021secure} that captures the enrolled user's profile. During the authentication stage, newly received signals are compared against the enrolled profile to determine the legitimacy of the user.

\begin{figure}[t]
  \centering
  \begin{minipage}[t]{0.48\linewidth}
    \centering
    \includegraphics[width=\linewidth]{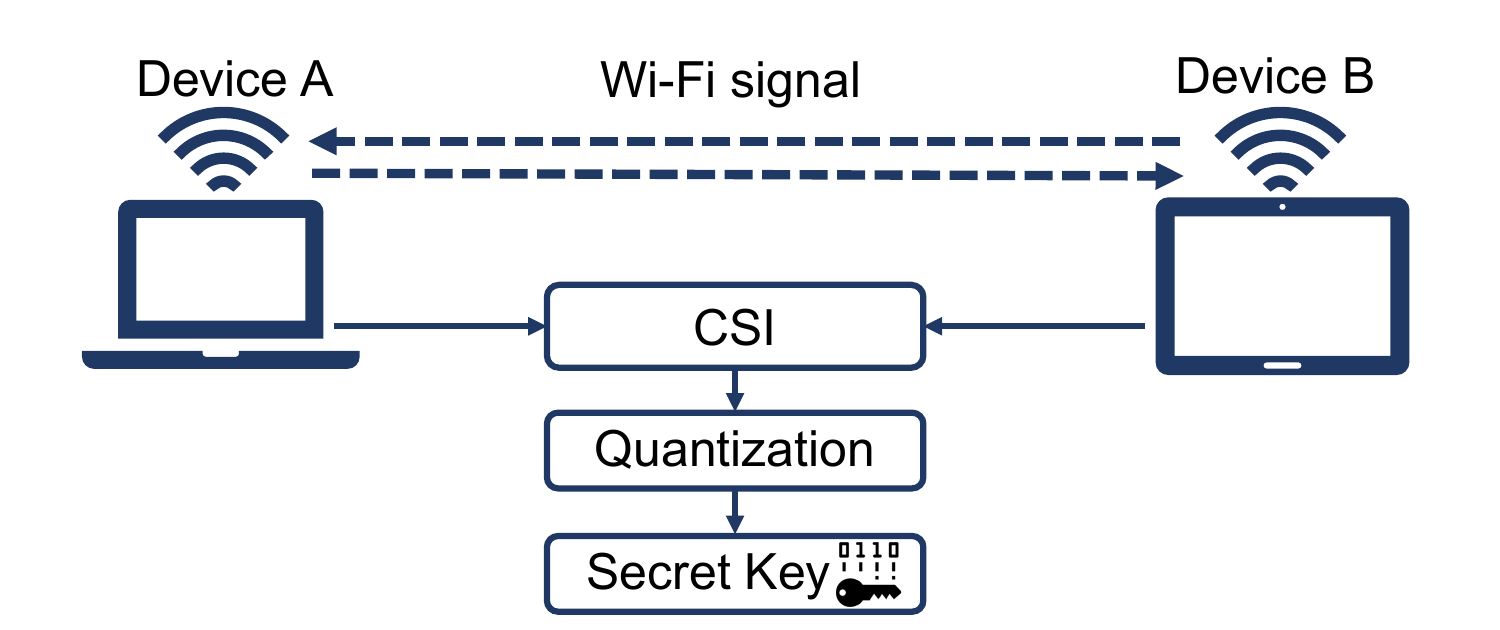}
    \captionof{figure}{Wi-Fi-based secret key generation.}
    \label{fig:secret-key}
  \end{minipage}
  \hfill
  \begin{minipage}[t]{0.48\linewidth}
    \centering
    \includegraphics[width=\linewidth]{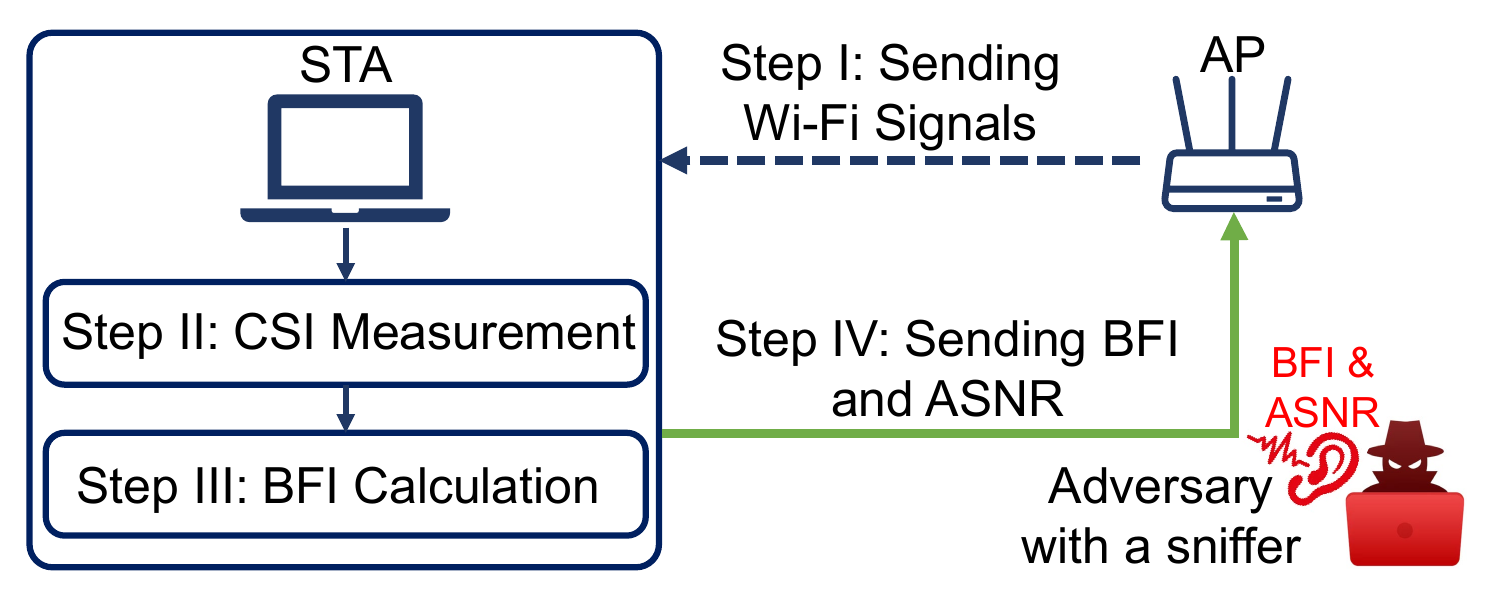}
    \captionof{figure}{Wi-Fi sounding procedure. When the STA sends the BFI and ASNR back to the AP, an adversary can easily capture the BFI and ASNR using a passive sniffer.}
    \label{fig:BFI}
  \end{minipage}
\end{figure}

\textbf{Secret Key Generation.}
Figure~\ref{fig:secret-key} shows an overview of Wi-Fi-based secret key generation. Specifically, two devices exchange a Wi-Fi packet within a short time interval, during which the wireless channel can be considered reciprocal. As a result, both devices experience the same channel conditions and independently extract the shared CSI from the received packets. Such shared CSI is then quantized into binary sequences to generate a shared secret key for both devices~\cite{liu2013fast, peng2017secret, ji2021wireless}.



\subsection{BFI Basics}

BFI is a type of Wi-Fi information defined in the IEEE 802.11ac/ax standard~\cite{6687187, 9442429} that characterizes the wireless channel to facilitate Multi-User Multiple-Input Multiple-Output (MU-MIMO) communication. With BFI, the Wi-Fi AP can adjust the complex weights of the transmitted signals across its antennas, thereby enhancing the signal reception at the receiving STA through beamforming. Because the AP only requires the information to compute weights and to minimize transmission overhead, the BFI is designed as a partial and compressed representation of the CSI. Assuming a scenario with one AP and one STA, the Wi-Fi sounding procedure is shown in Figure~\ref{fig:BFI}, which includes the following steps.

\textbf{Step I: Sending Wi-Fi Signals.} 
The AP initiates the procedure by broadcasting a Null Data Packet Announcement (NDPA) to the STA, followed by a Null Data Packet (NDP) for CSI measurement.

\textbf{Step II: CSI Measurement.}
We assume that the AP is equipped with $M$ antennas, while the STA is equipped with $N$ antennas. The STA estimates the CSI measurement, which can be represented as $H_k \in C^{N\times M}$, where $k$ denotes the subcarrier index.

\begin{figure*}[t]
  \centering
  \includegraphics[width=\linewidth]{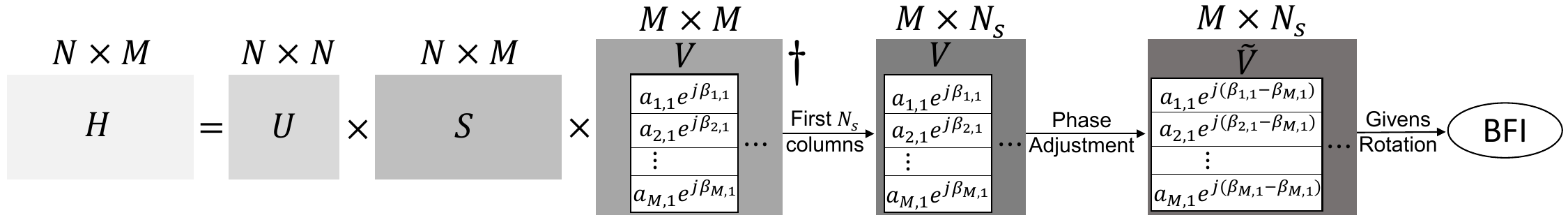}
  \caption{BFI calculation process.}
  \label{fig:svd-bfi}
\end{figure*}

\textbf{Step III: BFI Calculation.}
The BFI calculation process is illustrated in Figure~\ref{fig:svd-bfi}.
We first perform SVD on CSI:
\begin{equation} \label{eq:1}
    H_k = U_k S_k V_k^\dagger,
\end{equation}
where $U_k$ and $V_k$ are unitary matrices, $S_k$ is a diagonal matrix with nonnegative singular values $\sigma$, and $(\cdot)^{\dagger}$ denotes the Hermitian transpose. The right matrix $V_k$ determines the beamforming directions. 
For each element $(m_1,m_2)$ in $V_k$ ($m_1,m_2 \in \{1,2,\ldots,M\}$), the element can be expressed as:
\begin{equation} 
[V_k]_{m_1,m_2} = a_{m_1,m_2} e^{j\beta_{m_1,m_2}},
\label{eq:bfi}
\end{equation}
where $a$ and $\beta$ denote the amplitude and phase of $V_k$, respectively. According to the IEEE 802.11ac/ax standard, only the first $N_s$ columns of the right singular matrix $V_k$ are retained and subsequently compressed for feedback, where $N_s$ denotes the number of spatial streams and satisfies $N_s \le \min(M, N)$.

Next, the phase of all elements in each column is adjusted by subtracting the phase of the last-row element. The adjusted matrix $\tilde{V}_k$ is given by:
\begin{equation} 
[\tilde V_k]_{m_1,m_2} = a_{m_1,m_2} e^{j(\beta_{m_1,m_2}-\beta_{M,m_2})}.
\end{equation}

After the adjustment, the last row of each column has a phase of zero. This operation preserves the relative phase relationships within the matrix while discarding original phase. 

Then, the matrix $\tilde V_k$ is further decomposed to enable a compact feedback representation. Specifically, $\tilde V_k$ is written as a product of diagonal phase matrix $D_{k,i}$ and Givens rotation matrix $G_{k,\ell,i}$, which are defined as:
\begin{equation}
    D_{k,i} = \mathrm{diag}\!\left(I_{i-1}, e^{j\phi_{k,i,i}}, \ldots, e^{j\phi_{k,M-1,i}}, 1\right),
\end{equation}
\begin{equation}
    G_{k,\ell,i} =
    \begin{bmatrix}
    I_{i-1} & & & & \\
    & \cos\psi_{k,\ell,i} & 0 & \sin\psi_{k,\ell,i} & \\
    & 0 & I_{\ell-i-1} & 0 & \\
    & -\sin\psi_{k,\ell,i} & 0 & \cos\psi_{k,\ell,i} & \\
    & & & & I_{M-\ell}
    \end{bmatrix},
\end{equation}
where $i$ denotes the column index of $\tilde V_k$ with $1 \le i \le N_s$, $I$ is the identity matrix, and $\ell$ is the rotation index satisfying $i+1 \le \ell \le M$. 
The angles $\phi_{k,i}$ and $\psi_{k,\ell,i}$ are quantization parameters defined by the IEEE 802.11ac/ax standard.
We can directly extract the angles using a sniffer. By cascading these operations, we obtain:
\begin{equation}
    \tilde V_k =
    \left(
    \prod_{i=1}^{\min(N_s, M-1)} D_{k,i}
    \prod_{\ell=i+1}^{M} G_{k,\ell,i}^{\mathcal{T}}
    \right) I_{M\times N_s},
    \label{givens_rotation}
\end{equation}
which represents the BFI and is fully characterized by the angles $\phi_{k,i}$ and $\psi_{k,\ell,i}$. $(\cdot)^{\mathcal{T}}$ denotes the transpose operation. Note that deriving BFI from CSI via SVD is an irreversible transformation~\cite{wu2023enabling}.

\textbf{Step IV: Sending BFI and ASNR.} 
The STA then sends the BFI back to the AP. According to the IEEE 802.11ac/ax standard, BFI is transmitted in plaintext without encryption. To indicate the quality of the selected beams, the STA also sends the average signal-to-noise ratio (ASNR) over all subcarriers to the AP, which can be derived from the singular values.
Let the SVD of subcarrier $k$ be $ H_k= U_kS_kV_k^{\dagger}$ with 
$ S_k={diag}(\sigma_{k,1},\ldots,\sigma_{k,N_s})$.
The ASNR $\Upsilon$ of stream $i$ (in dB), averaged across $K$ subcarriers, can be expressed as:
\begin{equation}
  \Upsilon_i \;=\; \frac{1}{K}\sum_{k=1}^{K} 10\log_{10}\!\left(\frac{P_{TX}\,\sigma_{k,i}^{2}}{P_{N}}\right),
  \qquad i=1,\ldots,N_s ,
  \label{eq:asnr_def}
\end{equation}
where $P_{TX}$ denotes the transmit power and $P_N$ is the measured noise power.
The mean of the singular values of stream $i$ across all $K$ subcarriers can be denoted as:
\begin{equation}
  \bar{\sigma}_{i}
  = \left(\prod_{k=1}^{K}\sigma_{k,i}\right)^{\!1/K}
  = 10^{\,\Upsilon_i/20}\sqrt{\frac{P_{N}}{P_{TX}}},
  \qquad i=1,\ldots,N_s .
  \label{eq:asnr_geom_mean}
\end{equation}

During this transmission, an adversary can easily capture the BFI and ASNR using a passive sniffer as shown in Figure~\ref{fig:BFI}, which can be implemented with a commodity Wi-Fi device~\cite{haque2023wi}.
It is worth noting that the only type of BFI accessible to the adversary is the downlink (DL) BFI, which is transmitted from the STA to the AP. In this work, BFI refers to DL BFI by default, unless otherwise specified.

\section{Attack Design}

In this section, we first introduce the threat model and the attack overview. We then present BFIAttack in both single-antenna STA and multi-antenna STA scenarios.

\subsection{Threat Model}

\textbf{Attack Goals.}
Our attack is a targeted attack. In the context of device or user authentication, the adversary's goal is to impersonate a specific legitimate device or user within the wireless system. For secret key generation, the adversary seeks to crack or infer the shared secret key between devices.


\textbf{Adversary Capabilities.}
To achieve these goals, we assume the adversary has the knowledge of BFI and utilizes a commodity Wi-Fi device (e.g., a laptop) as a passive sniffer to extract the BFI from over-the-air transmissions.
The device can also collect the CSI between the adversary and the STA. 
The adversary has neither access to the legitimate user's/device's CSI nor control over any benign devices.
The adversary can infer the type of physical-layer security application in use and detect the beginning of the security application by analyzing passive Wi-Fi traffic patterns. For example, short, one-time burst packets typically indicate device authentication; continuous streams of packets suggest user authentication; and bursts of bidirectional probe packets are characteristic of secret key generation. The adversary can observe the environment surrounding the AP and STA and infer the distance between the AP and the STA. The adversary also knows public information about AP and STA hardware (e.g., bandwidth and number of antennas).

For device or user authentication, the adversary does not have the knowledge of the authentication model employed by the verifier. Nevertheless, the adversary can precode signals using the reconstructed CSI and transmit them to the verifier, ensuring that the resulting CSI observed by the verifier is identical to the reconstructed CSI.
This precoding and transmission can be implemented using software-defined radio (SDR)~\cite{ulversoy2010software, adib2013see}.
For secret key generation, we assume that the key generation algorithm is public and known to the adversary. This assumption is consistent with Kerckhoffs's principle~\cite{wang2018physical, stinson2005cryptography}, which asserts that secure systems should not rely on obscurity. 
Each instance in which the adversary precodes and transmits signals to the verifier to impersonate the legitimate device/user, or attempts to infer the secret key, is referred to as one attack attempt. The adversary is constrained to a limited number of attack attempts (e.g., no more than 5 attack attempts).


\subsection{Attack Overview}

\begin{figure*}[t]
    \centering
    \includegraphics[width=\linewidth]{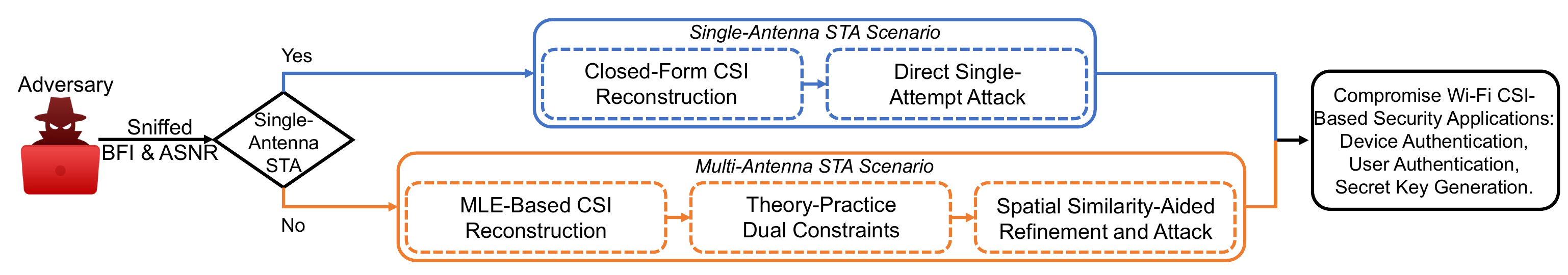}
    \caption{
    Attack overview.
    }
    \label{fig:overview}
\end{figure*}

The core idea of BFIAttack is to leverage passively sniffed BFI to reconstruct the legitimate user's or device's CSI, thereby compromising Wi-Fi CSI-based physical-layer security applications, such as impersonating the legitimate user or device, or inferring the secret key.
The attack overview is shown in Figure~\ref{fig:overview}. 

The adversary captures BFI and ASNR by passively sniffing Wi-Fi packets during the execution of authentication or key generation by the legitimate user/device, as shown in Figure~\ref{fig:BFI}. Next, the adversary determines whether the STA is single-antenna or multi-antenna. This can be done by inspecting publicly available hardware information of the STA or by analyzing the sniffed packets.

In the single-antenna STA scenario, the adversary reconstructs the CSI using a closed-form solution derived from the mathematical relationship between BFI and CSI. We note that this reconstruction yields only one reconstructed CSI for each BFI, due to the deterministic nature of the closed-form solution. Subsequently, the adversary only conducts a single-attempt attack within one attack trial, wherein the reconstructed CSI is directly used to precode signals in order to compromise device or user authentication or infer the secret key.

In the multi-antenna STA scenario, the adversary employs an MLE-based CSI reconstruction because no closed-form solution exists. The resulting MLE problem is non-convex and admits multiple solutions, producing several reconstructed CSIs for each BFI. Therefore, the adversary applies theory-practice dual constraints to filter infeasible CSIs. Remaining CSIs are then fine-tuned using a spatial similarity-aided refinement that exploits correlations among antenna pairs. Finally, multiple refined CSIs are used to precode signals or to infer the secret key. In this scenario, the adversary can conduct multiple attack attempts within one attack trial.



It is worth noting that CSI is a complex-valued number comprising both amplitude and phase components~\cite{ren2020liquid, xie2015precise}. However, many Wi-Fi-based security applications primarily utilize the CSI amplitude or zero-weight phase~\cite{pei2014channel, gu2021secure, xin2016freesense, liu2013fast, peng2017secret, ji2021wireless, wang2016csideepfi, song2021enhancing, chen2021authenticating, jing2023multi}, as the CSI phase is often contaminated by various random distortions~\cite{qu2024guessing}, including random initial phase, sampling frequency offset, and central frequency offset. Accordingly, in this work, we mainly focus on reconstructing the CSI amplitude in the attack design.

\subsection{Attack on Single-Antenna STA Scenario}


Existing work, BFMSense~\cite{yi2024bfmsense}, has shown that a mathematical relationship exists between BFI and CSI that can facilitate Wi-Fi sensing applications. In this work, we further demonstrate that this relationship can be exploited to reconstruct CSI in a closed-form manner, thereby enabling effective attacks on Wi-Fi-based physical-layer security systems.

\textbf{Closed-Form CSI Reconstruction.}
For each AP antenna index $u$ and STA antenna index $v$, the CSI can be expressed as:
\begin{equation}
    H_{v,u}=b_{v,u} e^{j\theta_{v,u}},
    \label{eq:csi}
\end{equation}
where $b$ and $\theta$ are the amplitude and phase of CSI. We then construct the equation~\cite{yi2024bfmsense}:
\begin{equation}
    {H}^\dagger{H}=VS^\dagger U^\dagger USV^\dagger=VS^\dagger SV^\dagger.
\end{equation}
Considering Equation~\eqref{eq:csi} and Equation~\eqref{eq:bfi}, the $(m_1,m_2)$ element of ${H}^\dagger{H}$ and $VS^\dagger SV^\dagger$ can be expressed as:
\begin{equation}
[{H}^\dagger{H}]_{m_1,m_2}
= \sum_{n=1}^N b_{n,m_1}b_{n,m_2}\,e^{j(\theta_{n,m_2}-\theta_{n,m_1})},
\label{eq:9}
\end{equation}
\begin{equation}
    [VS^\dagger SV^\dagger]_{m_1,m_2}=\sum_{n=1}^N \sigma_n^{2}\,a_{m_1,n}a_{m_2,n}\,e^{j(\beta_{m_1,n}-\beta_{m_2,n})}.
    \label{eq:10}
\end{equation}


It is important to note that only in the single-antenna STA scenario (i.e., if and only if $N = 1$), the summation notations in Equation~\eqref{eq:9} and Equation~\eqref{eq:10} can be eliminated, and a closed-form solution for CSI reconstruction can be derived: 
\begin{equation}
    b_{1,m_1}b_{1,m_2}\,e^{j(\theta_{1,m_2}-\theta_{1,m_1})}=\sigma_1^{2}\,a_{m_1,1}a_{m_2,1}\,e^{j(\beta_{m_1,1}-\beta_{m_2,1})},
\end{equation}
where $\sigma_1$ is the $(1,1)$ element in the diagonal matrix $S$, which is then substituted with the captured ASNR.



To calculate the CSI amplitude on each AP antenna, we let $m_1=m_2$. Therefore, we can obtain:
\begin{equation}
    b_{1,m_1}^{2} = a_{m_1,1}^{2} \cdot \sigma_1^{2}.
\end{equation}
Thus, the closed-form expression for CSI amplitude can be written as: 
\begin{equation}
    b_{1,m_1} = a_{m_1,1} \cdot \sigma_1.
\end{equation}


Although we primarily focus on reconstructing CSI amplitude, we can let $m_1\neq m_2$ and obtain the closed-form expression for the relative CSI phase between AP antennas:
\begin{equation} \label{equ14}
\quad
    \theta_{1,m_1}-\theta_{1,m_2}=\beta_{m_2,1}-\beta_{m_1,1}.
 \end{equation}

\begin{figure}[t]
  \centering
  \begin{subfigure}[t]{0.38\linewidth}
    \centering
    \includegraphics[width=\linewidth]{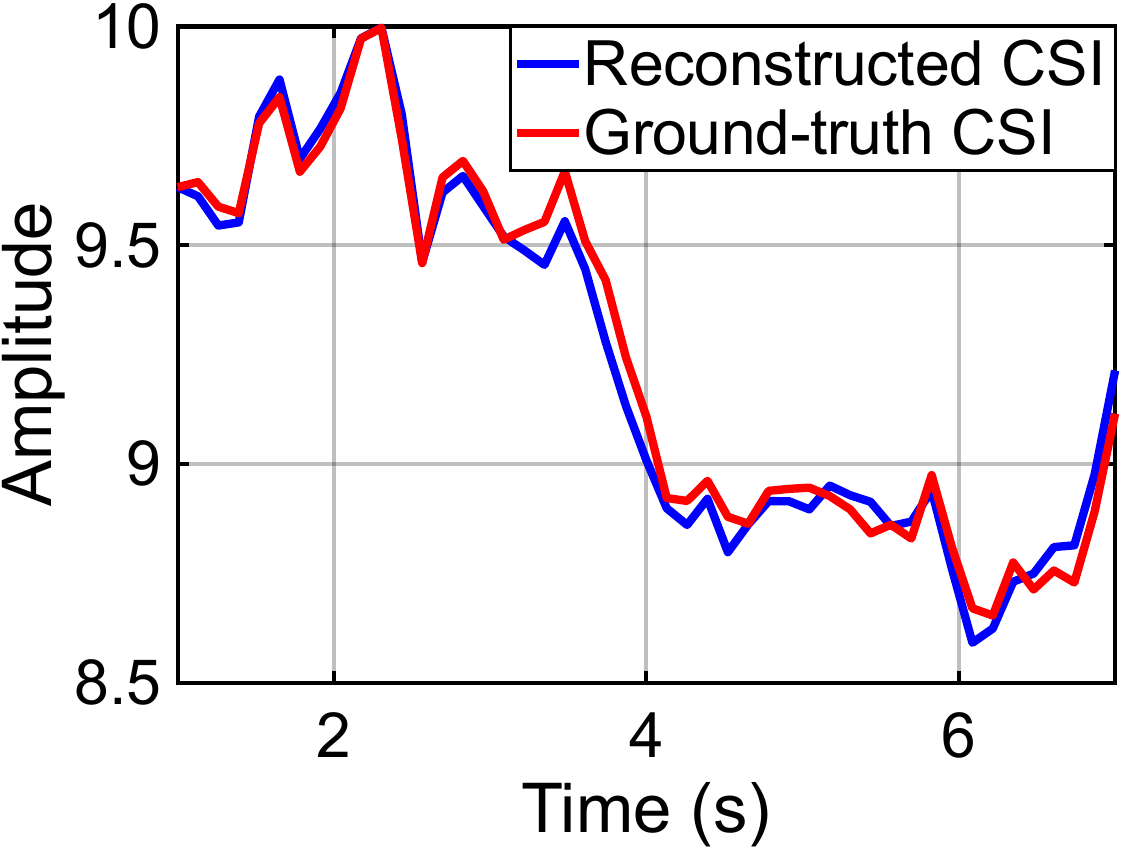}
    \caption{CSI amplitude over time.}
    \label{fig:a}
  \end{subfigure}\hspace{0.05\linewidth}
  \begin{subfigure}[t]{0.38\linewidth}
    \centering
    \raisebox{2pt}{\includegraphics[width=\linewidth]{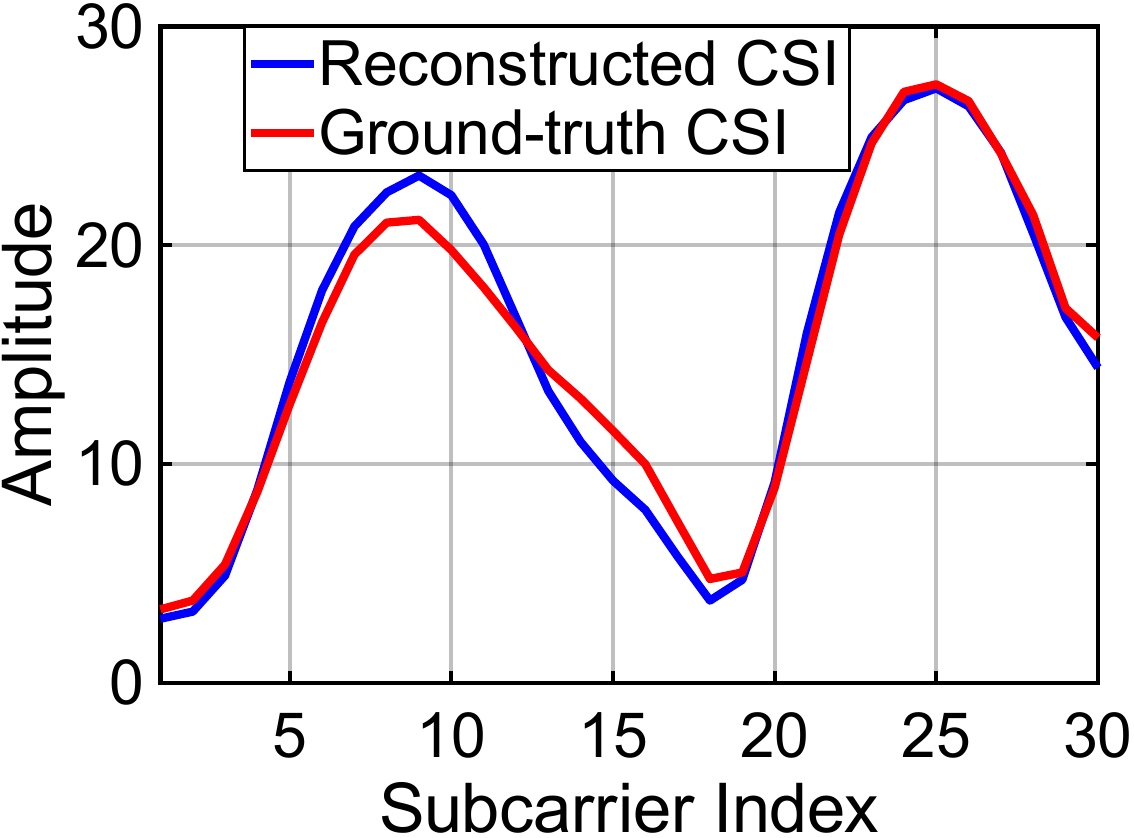}}
    \caption{CSI amplitude over subcarrier index.}
    \label{fig:b}
  \end{subfigure}
  \caption{Examples for CSI reconstruction in the single-antenna STA scenario.}
  \label{fig:single_example}
\end{figure}

We validate CSI reconstruction through preliminary experiments that simultaneously capture ground-truth CSI (i.e., a user's CSI) and BFI from the same Wi-Fi devices to make sure they represent the same channel state. Figure \ref{fig:single_example} shows examples for CSI reconstruction in the single-antenna STA scenario. 
Our results show that the reconstructed CSI closely aligns with the ground-truth CSI (i.e., user's or device's CSI) in both the time and frequency (i.e., subcarrier) domains. 
We can still observe subtle differences between the reconstructed CSI and the ground-truth CSI. This is because the adversary can obtain only the average ASNR, which is an average of singular values across subcarriers, rather than the exact singular values on each subcarrier. 
Nevertheless, this strong correspondence demonstrates the effectiveness of our method for attacking Wi-Fi-based physical-layer security applications.

\textbf{Direct Single-Attempt Attack.}
Given one BFI, only one corresponding CSI can be reconstructed due to the deterministic nature of such closed-form reconstruction. Therefore, in the single-antenna STA scenario, the number of attack attempts is inherently limited to one within each attack trial. Consequently, the adversary can directly leverage the reconstructed CSI to perform attacks. The complete attack procedure is as follows:

\begin{enumerate}[label=(\roman*)] 
\item The adversary passively obtains both the BFI and the ASNR by sniffing Wi-Fi packets transmitted during the legitimate user's or device's execution of authentication or secret key generation between devices.
  \item The adversary uses the closed-form method to reconstruct CSI amplitude for each antenna and subcarrier.
  \item The adversary precodes signals using the reconstructed CSI and sends them to the verifier (i.e., STA) to impersonate the legitimate user or device. For secret key generation, the adversary inputs the reconstructed CSI into the same key generation algorithm used by the user to recover the shared secret key.
\end{enumerate}

\subsection{Attack on Multi-Antenna STA Scenario}

The multi-antenna STA scenario introduces more challenges for conducting attacks on Wi-Fi-based physical-layer security. 
First, a multi-antenna STA scenario cannot be decomposed into $N$ independent single-antenna STA scenarios to apply the closed-form solution.
The reason is that performing SVD on an $M \times N$ CSI matrix ($N>1$) yields fundamentally different singular values compared to those from an $M \times 1$ matrix, even when the matrix elements are identical. This arises because SVD captures the global structure of the matrix rather than treating each part independently.
Second, the only existing approach for reconstructing CSI from BFI is BeamSense~\cite{wu2023enabling}, which requires bidirectional BFI (i.e., both downlink (DL) and uplink (UL)). However, fewer than $1\%$ of commodity Wi-Fi devices support bidirectional BFI extraction~\cite{wu2023enabling}, which severely limits its practicality and real-world attack surface.
In contrast, we aim to achieve attacks using only unidirectional BFI (i.e., downlink (DL) BFI), which is supported by the vast majority of commodity Wi-Fi devices.

\textbf{MLE-Based CSI Reconstruction Using DL BFI.}
We adopt an MLE-based approach to estimate channel parameters and reconstruct CSI using only the DL BFI. 
To analyze the relationship between the sniffed DL BFI and an unknown UL matrix, we begin from the decomposition of CSI with both DL and UL:
\begin{equation}
    H = U_{DL} S V_{DL}^\dagger
    = {\overline V}_{UL} S U_{UL}^{\mathcal{T}},
    \label{eq:bidirectional}
\end{equation}
where $\overline{(\cdot)}$ denotes the complex conjugation.
In BFI, we can obtain only $\tilde V_{DL}$, which is the phase-adjusted version of $V_{DL}$ defined by $ V_{DL}=\tilde V_{DL} D_{DL}$, where $D_{DL}$ is a diagonal phase matrix.
Similarly, we have $V_{UL}=\tilde V_{UL} D_{UL}$. 
According to Equation~\eqref{eq:bidirectional}, there must exist another diagonal phase matrix $D_s$ such that $U_{DL} = \overline{V}_{UL} D_s$.
Therefore, we obtain:
\begin{equation}
    H = \overline {\tilde V}_{UL} \overline D_{UL} D_s S D_{DL}^{\dagger} \tilde V_{DL}^{\dagger}.
    \label{eq:conbine_d}
\end{equation}
Since $D_{UL}$, $D_{DL}$, and $D_s$ are all diagonal phase matrices, we define $D'=\overline D_{UL} D_s D_{DL}^{\dagger}$, where $D'$ is also a diagonal phase matrix. 
Hence, $H$ can be rewritten as:
\begin{equation}
    H = \overline {\tilde V}_{UL} D' S\,\tilde V_{DL}^{\dagger}.
    \label{eq:conbine_d}
\end{equation}
By multiplying both sides by $\tilde V_{UL}^{\mathcal{T}}$ on the left and $\tilde V_{DL}$ on the right, we have:
\begin{equation}
    \tilde V_{UL}^{\mathcal{T}} H\tilde V_{DL} = D' S.
    \label{eq:construct}
\end{equation}

We then build a matrix $T$:
\begin{equation}
    T
    = [\tilde{V}_{UL}(\{\phi_{UL,i},\psi_{UL,\ell,i}\})]^\mathcal{T}
       {H}(\Omega)
      \tilde{V}_{DL},
      \label{eq:los}
\end{equation}
where $\tilde V_{UL}$ is obtained from  Equation~\eqref{givens_rotation} using
$(\phi_{\mathrm{UL},i},\,\psi_{\mathrm{UL},\ell,i})$, and $H$ is computed based on the channel parameter set $\Omega$: 
\begin{equation}
    \Omega = \{\alpha_p,t_p,\chi_p,\gamma_p,
              \phi_{UL,i},\psi_{UL,\ell,i}\},
\end{equation}
where $p$ is the index of propagation paths, $\alpha$ denotes the attenuation, $t$ is the delay introduced by time of flight (ToF) and Doppler frequency shifts, $\chi$ is the angle of arrival (AoA), and $\gamma$ is the angle of departure (AoD)~\cite{ren2022gopose, ren2021winect}. 

In theory, the matrix $T$ should be equal to the matrix $D'S$, and both are diagonal matrices. 
However, since the diagonal phase matrix $D'$ is unknown, it is not feasible to directly match $T$ with the complex-valued matrix $D'S$. 
Nevertheless, we can align their amplitudes as the amplitude of each diagonal entry in $D'S$ is preserved.
Specifically, for $D'S$, the amplitudes are equal to the singular values.
Therefore, our objective is twofold:
(1) the diagonal elements of $T$ should match the singular values $\sigma$, and
(2) the off-diagonal elements should be minimized, ideally approaching zero.
We achieve this objective by formulating an MLE problem that minimizes the following loss function:
\begin{equation}
    L(\Omega)
    = \sum_{i=1}^{N_s}\big(|T_{ii}|-\sigma_i)^2
      + \sum_{i=1}^{N_s}\sum_{\substack{j=1\\ j\ne i}}^{N_s}|T_{ij}|^2.
    \label{eq:opt_loss}
\end{equation}

Once the channel parameters are estimated, the CSI can be fully reconstructed as:
\begin{equation}
    H_k = \sum_{p=1}^P 
    \alpha_p \, e^{-j 2\pi f_k t_p}\;
    \mathbf{a}(f_k,\chi_p)\,\mathbf{d}(f_k,\gamma_p)^\dagger,
\end{equation}
where $p$ is the path index, $k$ is the subcarrier index, and $f_k$ denotes the frequency of subcarrier. For an antenna array with antenna spacing $s$, the transmitting steering vector $\mathbf{a}$ and receiving steering vector $\mathbf{d}$ are defined as:
\begin{equation}
    \mathbf{a}(f,\chi) =
    \big[\,1,\; e^{-j\frac{2\pi f}{c}s\sin\chi},\; \ldots,\;
    e^{-j\frac{2\pi f}{c}s(N-1)\sin\chi}\,\big],
\end{equation}
\begin{equation}
    \mathbf{d}(f,\gamma) =
    \big[\,1,\; e^{-j\frac{2\pi f}{c}s\sin\gamma},\; \ldots,\;
    e^{-j\frac{2\pi f}{c}s(M-1)\sin\gamma}\,\big],
\end{equation}
where $c$ is the speed of light.




Our MLE-based CSI reconstruction requires estimating multiple channel parameters, resulting in multidimensional search complexity. If an exhaustive search is employed, the computational complexity becomes prohibitive. 
Specifically, assuming $\mathcal{G}$ denotes the search space for each parameter, the total complexity is $\mathcal{G}^P_{\alpha} \times \mathcal{G}^P_{\chi} \times \mathcal{G}^P_{\gamma} \times \mathcal{G}^P_{t} \times \mathcal{G}_{\phi} \times \mathcal{G}_{\psi}$. 
To address this challenge, we adopt a coordinate descent method~\cite{wright2015coordinate} to iteratively approach maxima. In particular, we could replace the six-dimensional search with six one-dimensional searches. We can fix five (e.g., $\alpha, \chi, \gamma, t, \phi$) of the six parameters and search for the value of the remaining parameter (e.g., $\psi$) that yields a maximal output. Then, we repeat this process for the other parameters. As a result, we reduce the search space to $\sum_{p=1}^{P}\!\Bigl(\mathcal{G}_{\alpha} +\mathcal{G}_{\chi} +\mathcal{G}_{\gamma} +\mathcal{G}_{t}\Bigr) +\mathcal{G}_{\phi} +\mathcal{G}_{\psi}$, which dramatically reduces the computational complexity. As with most non-convex optimization problems, the performance of iterative search methods is highly sensitive to the choice of initialization points. Therefore, we adopt a strategy inspired by genetic search. We first perform optimization over a coarse-grained grid and select several optimal grid points as initialization candidates. A local search is then conducted from each of these points. Finally, the best result among these searches is selected. We repeat this process multiple times to produce multiple reconstructed CSIs for subsequent attacks.

\textbf{Theory-Practice Dual Constraints.}
We observe that the MLE problem in our work is inherently non-convex, leading the MLE-based CSI reconstruction to produce multiple sets of channel parameters that satisfy Equation~\eqref{eq:los}, thereby yielding multiple reconstructed CSIs.
Thus, we develop theory-practice dual constraints to filter infeasible reconstructed CSIs. Our key insight is that a reconstructed CSI resembling that of the legitimate user/device must simultaneously (1) satisfy the theoretical specifications defined in the IEEE 802.11ac/ax standard and (2) conform to real-world physical propagation characteristics, such as the distance between the AP and STA, which can be readily estimated by the adversary through passive observation.


To derive the theory constraint, we compute the upper bound and lower bound of the CSI amplitude limited by the IEEE 802.11ac/ax standard.  
According to Equation~\eqref{eq:conbine_d}, there is a matrix
\(D' = \mathrm{diag}(e^{j x_1}, \ldots, e^{j x_{N_s}})\)
which can represent the overall phase adjustment.  
\(D'\) introduces phase terms \(x_1,\ldots,x_{N_s}\), each ranging over \([0,\pi]\). 
Moreover, the uplink feedback matrix $V_{UL}$ is constrained by Givens rotation angles \(\phi \in [0,\pi]\) and \(\psi \in [0,\pi/2]\), as specified by the IEEE 802.11ac/ax standard.
$V_{DL}$ is a fixed matrix obtained directly from the passively captured BFI.
According to Equation~\eqref{eq:conbine_d}, each element in reconstructed CSI $H_{v,u}$ can be expressed as:
\begin{equation}
H_{v,u}
=
\sum_{i=1}^{N_s}
\sigma_i\,
[\tilde V_{UL}]_{v,i}\,
e^{j x_i}\,
[\tilde V_{DL}^{\dagger}]_{i,u},
\end{equation}
which is the sum of $N_s$ complex numbers, each of which is defined as:
\begin{equation}
z_i
=\sigma_i\,
[\tilde V_{UL}]_{v,i}\,
e^{j x_i}\,
[\tilde V_{DL}^{\dagger}]_{i,u}.
\end{equation}

When all complex numbers $z_i$ are phase-aligned, the $H_{v,u}$ achieves its maximum amplitude $|H_{max}|$:
\begin{equation}
    |H_{max}|=\sum_{i=1}^{N_s}\big|z_i\big|.
\end{equation}
This upper bound can be obtained by searching over the allowable ranges of the phase terms and Givens rotation angles in BFI specified in the IEEE 802.11ac/ax standard.


By the reverse triangle inequality, we have $|z_1|+|z_2|\ge |z_1+z_2|\ge \big||z_1|-|z_2|\big|$.
Accordingly, the following relation holds:
\begin{equation}
    \begin{aligned}
|z_1 + z_2 + \cdots + z_{N_s}|
&\ge \big|{{|z_{max}|}-{|\sum_{\substack {i=1 \\ i\ne max }}^{N_s} z_i|}}\big| \\
&\ge {|z_{max}|}-{|\sum_{\substack {i=1 \\ i\ne max }}^{N_s} z_i|}.
\end{aligned}
\label{eq:relation}
\end{equation}
By the triangle inequality $|z_1+z_2|\leq ||z_1|+|z_2|\big|$, we have:
\begin{equation}
    |\sum_{\substack {i=1 \\ i\ne max }}^{N_s} z_i| \leq \big|\sum_{\substack {i=1 \\ i\ne max }}^{N_s} |z_i|\big|.
\end{equation}
Equation~{\eqref{eq:relation}} can be rewritten as:
\begin{equation}
    \begin{aligned}
        |z_1 + z_2 + \cdots + z_{N_s}|
&\ge{|z_{max}|}-{|\sum_{\substack {i=1 \\ i\ne max }}^{N_s} z_i|} \\
&\ge{|z_{max}|}- {\big|\sum_{\substack {i=1 \\ i\ne max }}^{N_s} |z_i|\big|}\\
&=2|z_{\max}| - Z.
    \end{aligned}
\end{equation}

Then, the minimum amplitude of $H_{v,u}$, denoted by $|H_{\min}|$, is:
\begin{equation}
    |H_{\min}| = \max\!\big(0,\;2|z_{\max}| - Z\big),
    \label{eq:minimum}
\end{equation}
which represents the theoretical lower bound. 
Thus, the reconstructed CSI should satisfy $|H_{\min}| \le |{H}_{v,u}| \le |H_{\max}|$. Otherwise, it will be discarded.


To derive the practice constraint, we examine whether the ToF ($t_{\text{rec}}$) associated with the line-of-sight (LoS) path in the reconstructed CSI matches the ToF ($t_{\text{real}}$) of the LoS path between the AP and STA in the real world.
We note that $t_{\text{rec}}$ represents the pure ToF, as the Doppler shift of the LoS path is zero. Moreover, $t_{\text{rec}}$ is readily obtained during MLE-based CSI reconstruction.
The real-world ToF is $t_{\text{real}} = Dist_{AP-STA} / c$, where $Dist_{AP-STA}$ is the distance between the AP and STA, which can be passively observed by the adversary, and $c$ is the speed of light.
Considering that the ToF resolution of Wi-Fi signals is $\Delta t = 1/B$~\cite{xie2015precise}, where $B$ is the signal bandwidth, we check whether the reconstructed CSI satisfies: $t_{\text{real}} - \Delta t \le t_{\text{rec}} \le t_{\text{real}} + \Delta t$.
CSIs that do not satisfy this condition are discarded.




\textbf{Spatial Similarity-Aided Refinement and Attack.}
For the remaining reconstructed CSIs, we develop a spatial similarity-aided refinement method to enhance attack efficacy. In typical IoT and mobile devices, multiple antennas are close to each other, resulting in highly correlated wireless channels. Consequently, there exist strong correlations~\cite{mei2020envelope, nosrat2011mimo} among the CSI amplitudes across different AP-STA antenna pairs.

\begin{figure}[t]
  \centering
  \includegraphics[width=0.5\linewidth]{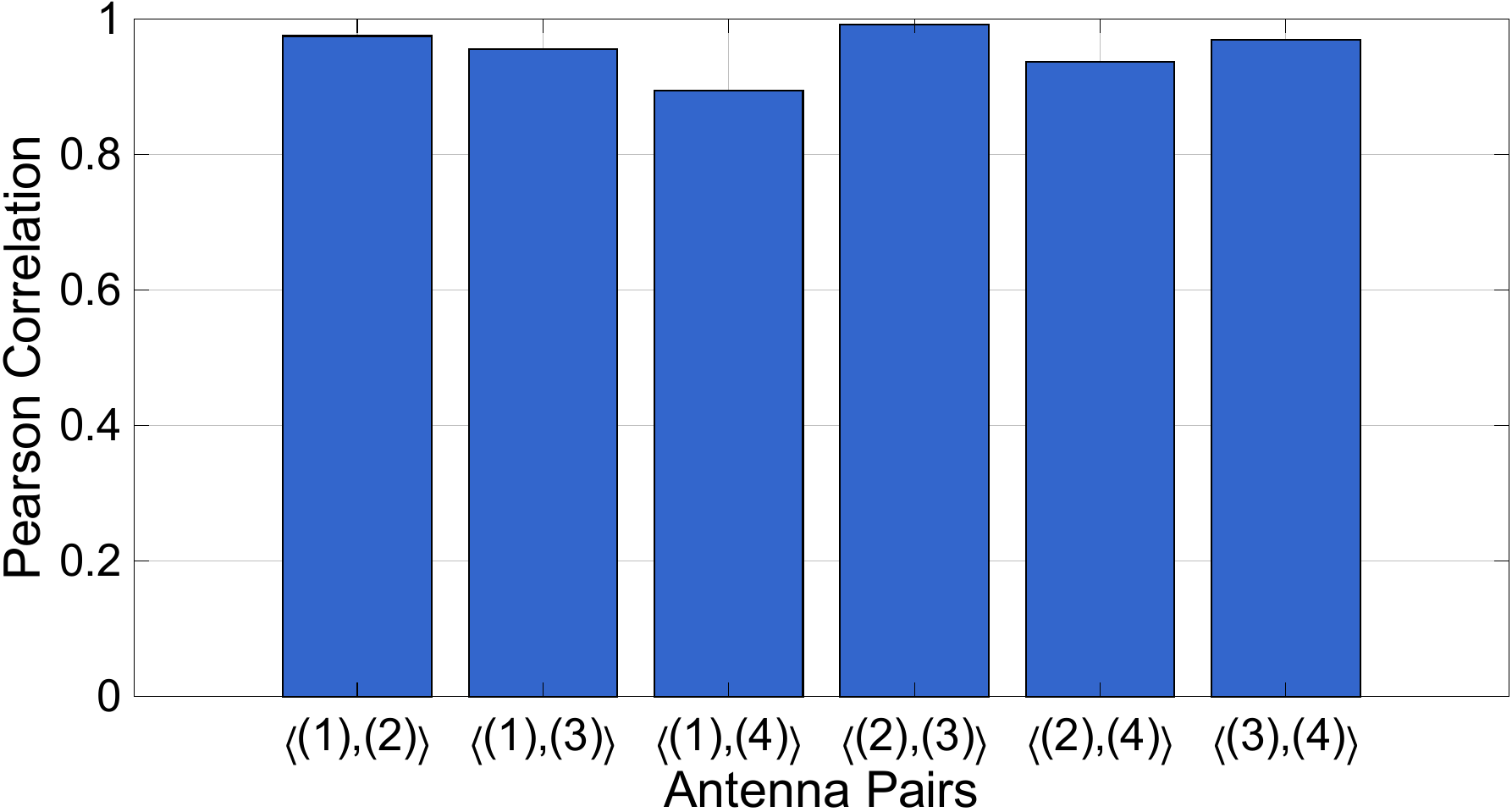} 
  \caption{The Pearson correlation coefficients across different antenna pairs.}
  \label{fig:pc}
\end{figure}

To validate the spatial similarity, we analyze real-world CSI amplitudes collected from a $2\times2$ multiple-input multiple-output (MIMO) system, where both the AP and the STA are equipped with two antennas. This configuration yields four AP-STA antenna pairs: (1) AP1-STA1, (2) AP1-STA2, (3) AP2-STA1, and (4) AP2-STA2. We collect CSI from 10 packets and compute the Pearson correlation coefficients between CSI amplitudes of the following six pairwise combinations:
$\langle(1),(2)\rangle$,
$\langle(1),(3)\rangle$,
$\langle(1),(4)\rangle$,
$\langle(2),(3)\rangle$,
$\langle(2),(4)\rangle$, and
$\langle(3),(4)\rangle$.
The Pearson correlation coefficients are shown in Figure~\ref{fig:pc}. We can observe strong correlations across antenna pairs. 

\begin{figure}[t]
  \centering
  \begin{subfigure}[b]{0.4\linewidth}
    \includegraphics[width=\linewidth]{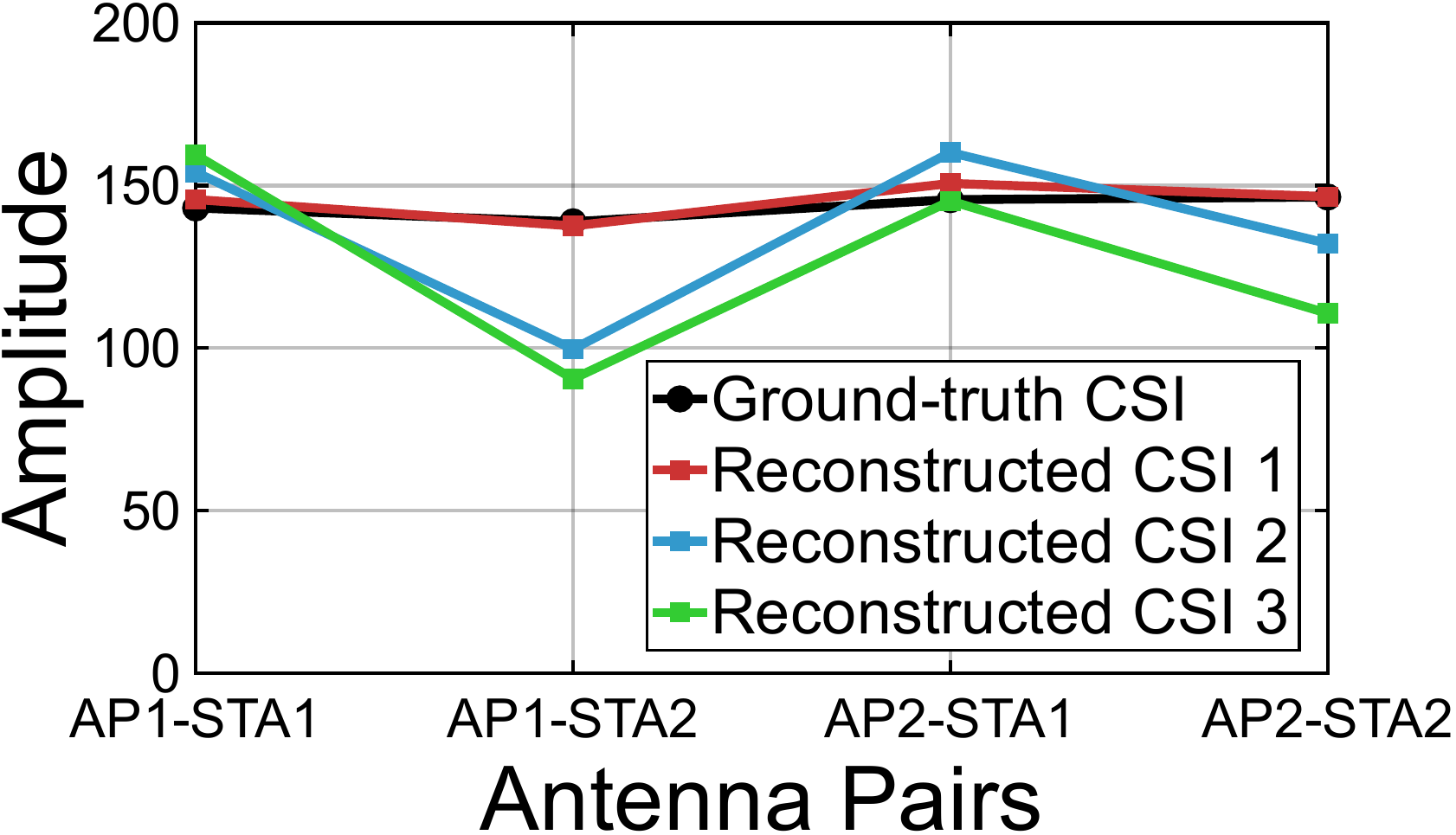}
    \caption{\mbox{Before refinement}.}
    \label{fig:a}
  \end{subfigure}\hspace{0.03\linewidth}
  \begin{subfigure}[b]{0.4\linewidth}
    \includegraphics[width=\linewidth]{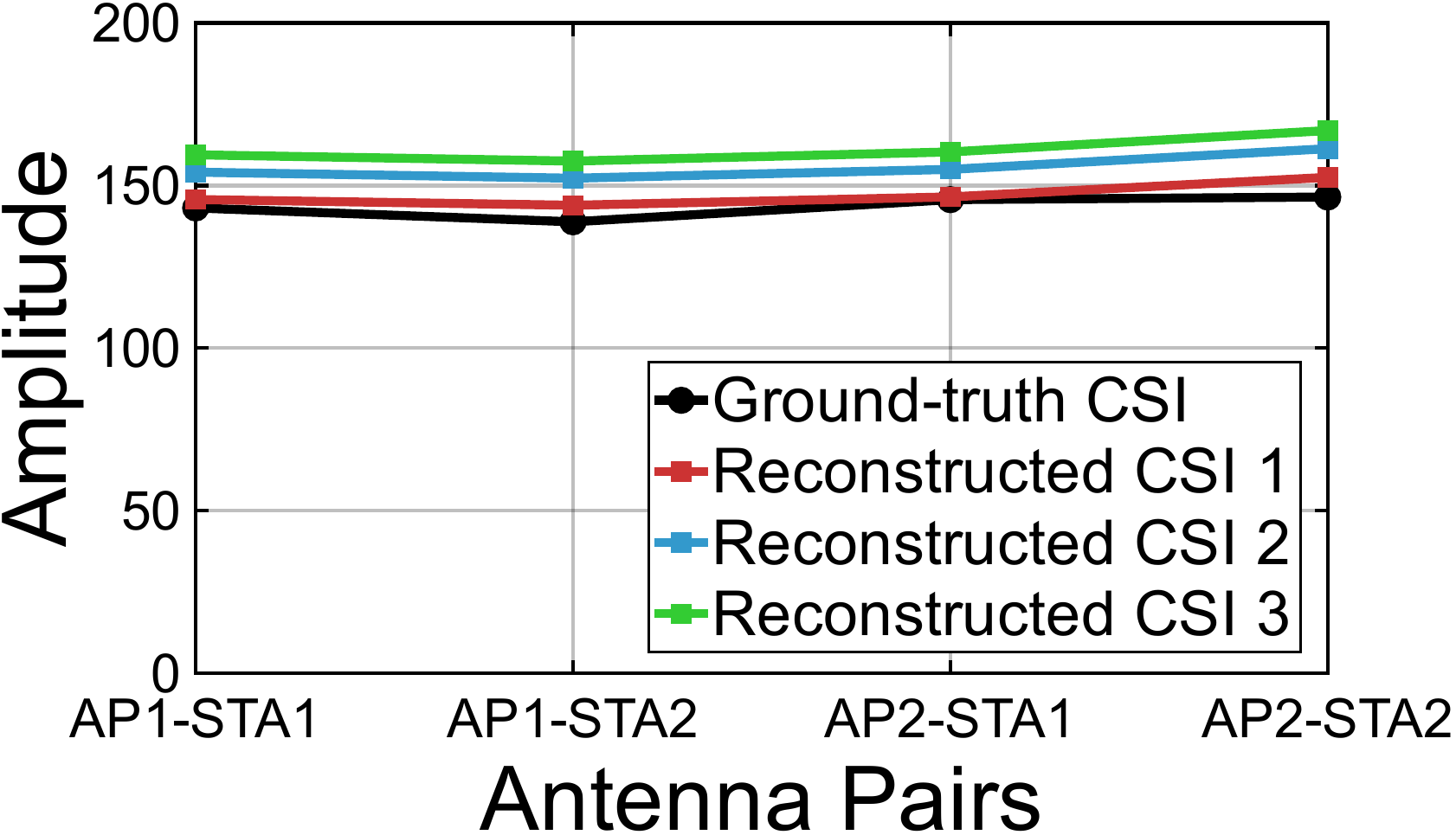}
    \caption{\mbox{After refinement}.}
    \label{fig:b}
  \end{subfigure}
  \caption{Spatial similarity-aided refinement for reconstructed CSIs.}
  \label{fig:dif_scale}
\end{figure}

Since strong the correlations can be approximated by a linear relationship~\cite{ratner2009correlation}, we employ a linear regression model~\cite{zou2003correlation} that uses the reconstructed CSI of one antenna pair to refine the reconstructed CSI of other antenna pairs. 
The adversary passively collects CSI values at various locations within the target environment from the STA and selects the antenna pair with the highest average correlation coefficient across all other pairs as the reference antenna pair. A linear regression model is then trained to characterize the relationships between the reference and non-reference antenna pairs. Specifically, for each non-reference antenna pair $\varepsilon \in [1, MN]$ ($\varepsilon \ne \varepsilon'$), the relationship is modeled as: $b_{\varepsilon} = \mu_{\varepsilon} \cdot b_{\varepsilon'}$, where $b_{\varepsilon}$ and $b_{\varepsilon'}$ denote the CSI amplitudes of antenna pair $\varepsilon$ and the reference pair $\varepsilon'$, respectively, and $\mu_{\varepsilon}$ is the weight for pair $\varepsilon$.
The model's training can be conducted offline. Once the weights are obtained, the adversary uses them to refine the reconstructed CSI for each antenna pair.
We collect CSI and BFI from the same device simultaneously and Figure~\ref{fig:dif_scale} illustrates the effectiveness of the proposed CSI refinement. Specifically, antenna pair (1) of the reconstructed CSI is selected as the reference, and the trained model is applied to adjust the CSIs of other antenna pairs. After refinement, the CSIs of all antenna pairs more closely align with the ground-truth CSI.

\begin{figure}[t]
  \centering
  \begin{subfigure}[t]{0.4\linewidth}
    \centering
    \includegraphics[width=\linewidth]{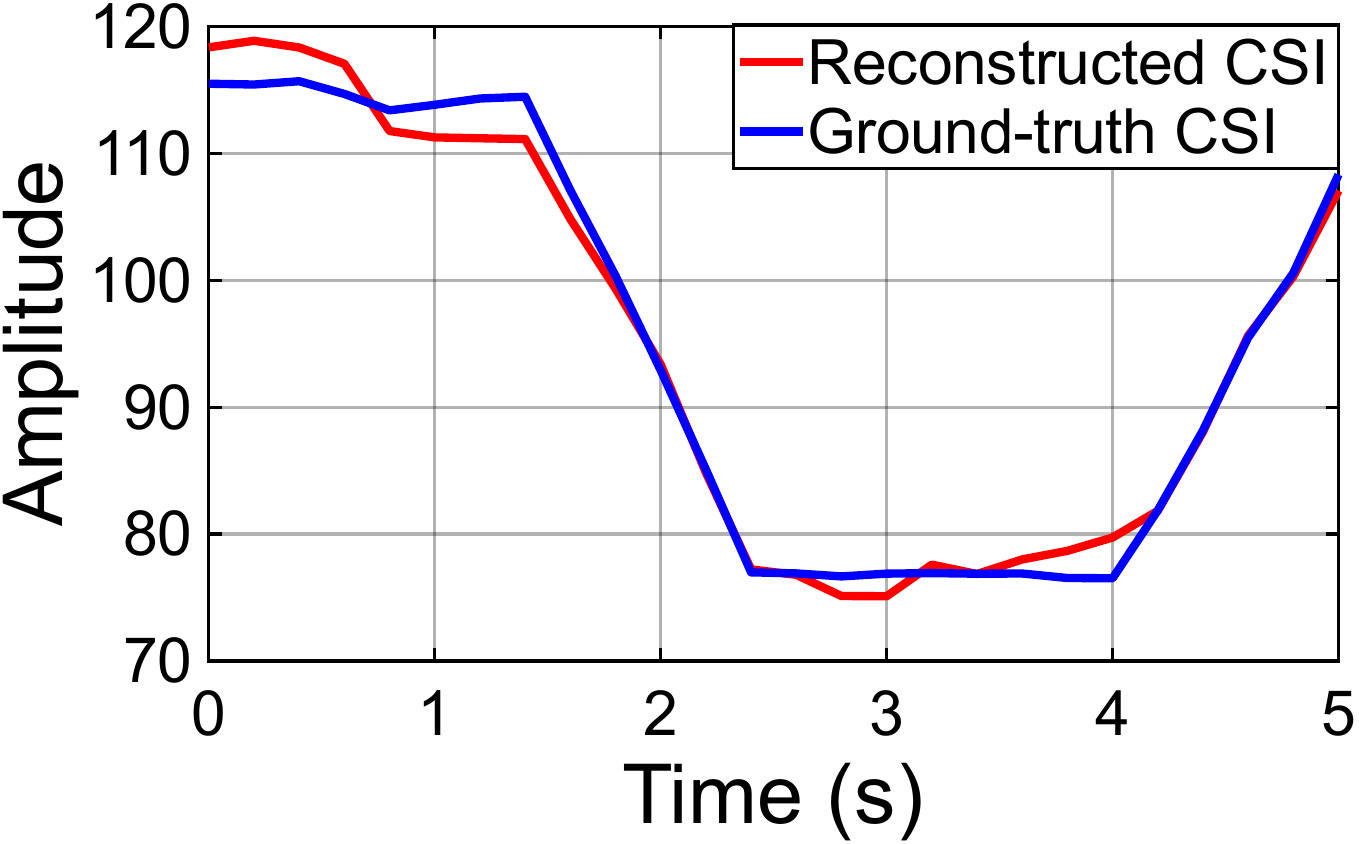}
    \caption{CSI amplitude over time.}
    \label{fig:a}
  \end{subfigure}\hspace{0.05\linewidth}
  \begin{subfigure}[t]{0.4\linewidth}
    \centering
    \raisebox{2pt}{\includegraphics[width=\linewidth]{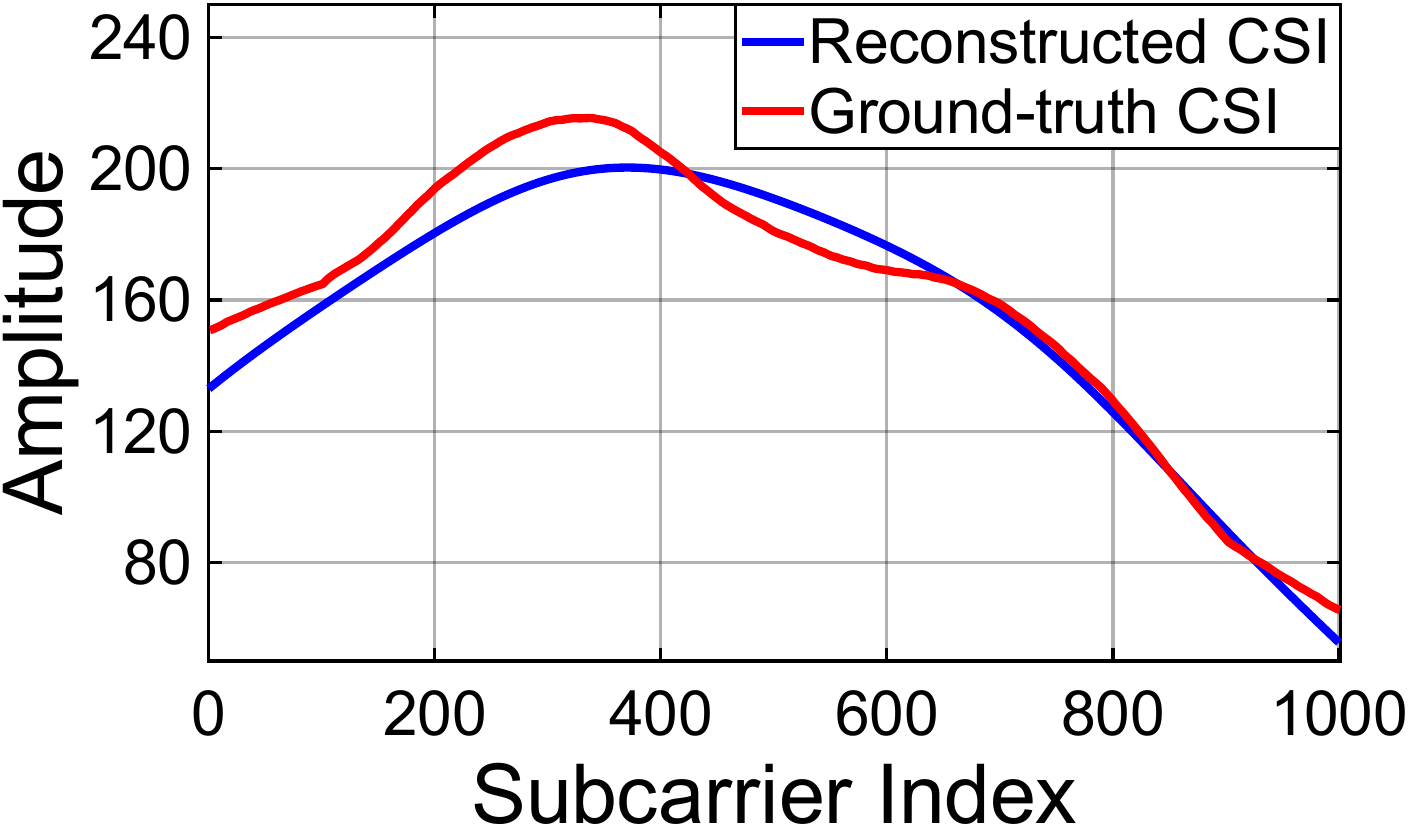}}
    \caption{CSI amplitude over subcarrier index.}
    \label{fig:b}
  \end{subfigure}
  \caption{\small Examples for CSI reconstruction in multi-antenna STA scenario.}
  \label{fig:scale}
\end{figure}

We further validate the overall performance of CSI reconstruction in the multi-antenna STA scenario. As shown in Figure~\ref{fig:scale}, the CSI reconstructed from BFI can match the ground-truth CSI in both the time and frequency (i.e., subcarrier) domains. 
In summary, the attack procedure for the multi-antenna STA is as follows: 

\begin{enumerate}[label=(\roman*)] 
    \item It is the same as in the single-antenna STA scenario.
    \item The adversary employs an MLE-based method to reconstruct the CSI amplitude for each antenna and subcarrier. Dual constraints are then applied to eliminate infeasible CSI candidates. The remaining CSIs are refined using a pre-trained linear regression model.
    \item The adversary uses each CSI to either precode signals for impersonation or to generate secret keys.
\end{enumerate}

\section{Evaluation}

\subsection{Experimental Setup}
\textbf{Devices.}
We evaluate BFIAttack using three types of commodity Wi-Fi routers configured as APs: the ASUS RT-AX86U, Linksys AC3000, and TP-Link AXE5400. A laptop equipped with an Intel AX210 network interface card (NIC) serves as the STA. Another laptop, also equipped with an Intel AX210 NIC, operates as a passive Wi-Fi sniffer and runs the BFI tool~\cite{haque2023wi} to extract BFI and ASNR. Both laptops can collect CSI using the PicoScenes CSI tool~\cite{jiang2021eliminating} for building Wi-Fi-based security applications or the offline linear regression model. 
The three APs are equipped with 4, 4, and 6 antennas, respectively, while both laptops have 2 antennas. 
To emulate the single-antenna STA scenario and evaluate the impact of the number of antennas, we selectively disable subsets of antennas on the AP and STA.
All devices support bandwidths of 20 MHz, 40 MHz, and 80 MHz. The default bandwidth is 80 MHz in this work.

\begin{figure}[t]
  \centering
  \begin{subfigure}[t]{0.25\linewidth}
    \centering
    \includegraphics[height=4cm]{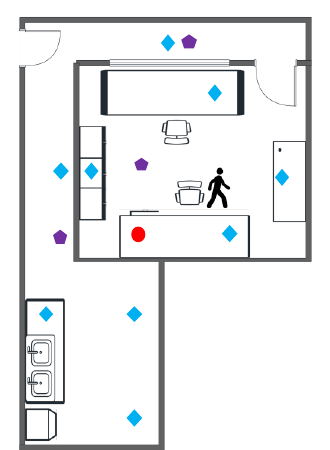}
    \caption{Laboratory.}
    \label{fig:lab}
  \end{subfigure}
  \hspace{0.005\linewidth}
  \begin{subfigure}[t]{0.25\linewidth}
    \centering
    \includegraphics[height=4cm]{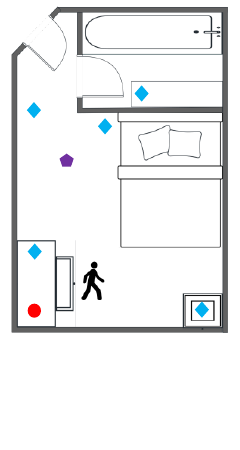}
    \caption{Apartment.}
    \label{fig:apt}
  \end{subfigure}
  \hspace{0.005\linewidth}
  \begin{subfigure}[t]{0.25\linewidth}
    \centering
    \includegraphics[height=4cm]{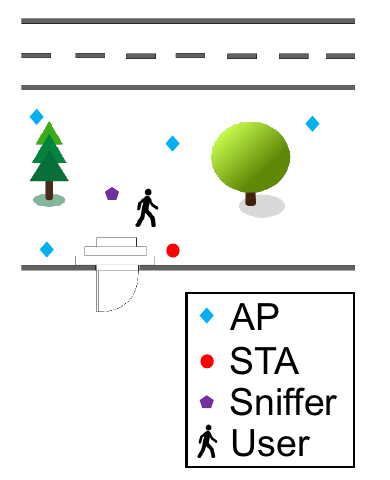}
    \caption{Outdoor.}
    \label{fig:out}
  \end{subfigure}
  \caption{Experimental environments.}
  \label{fig:setup}
\end{figure}

\textbf{Environments.}
We conduct experiments in three environments: a laboratory ($3.6m \times 3.5m$), an apartment ($3m \times 5m$), and an outdoor area ($4.8m \times 3m$), as shown in Figure~\ref{fig:setup}. The locations of the AP, STA, and sniffer are illustrated in Figure~\ref{fig:setup}. For device authentication, each user's device (i.e., AP) is placed at a distinct location for evaluation. For user authentication, the AP-STA pair closest to the user is selected. For secret-key generation, all AP-STA pairs are evaluated. We note that the environments encompass diverse conditions, including varying distances and non-line-of-sight (NLoS) conditions.


\textbf{Security Application Settings.}
For device authentication, we follow a general CSI-based framework~\cite{chen2021authenticating}, which authenticates the device using only one Wi-Fi packet. For each packet, we extract CSI amplitudes across all subcarriers and antenna pairs. We then train a machine learning classifier (e.g., SVM, CNN, or RNN) for authentication. A decision threshold is calibrated on the validation set to achieve a target false positive rate of 0.05.
For user authentication, we adopt gait-based approaches~\cite{zeng2016wiwho} that use multiple time-series CSI amplitudes to construct user profiles. Also, a machine learning classifier is trained as the authentication model, and a decision threshold is set to maintain a target false positive rate of 0.05.
For secret-key generation, we follow a publicly available algorithm~\cite{aldaghri2020physical}. Both the AP and STA independently extract CSI amplitude using one packet, apply denoising and normalization, and quantize CSI into bit sequences. Minor mismatches between the resulting sequences are corrected through a reconciliation process.

\textbf{Data Collection.}
For device authentication, we place 18 user devices (APs) at distinct locations and collect over 108,000 Wi-Fi packets. Unless otherwise specified, the default packet rate is 10 packets per second in this work.
For user authentication, five participants of varying heights, weights, and ages are recruited. Each participant walks for 5 seconds, repeated 180 times, yielding 450,000 Wi-Fi packets for gait-based authentication. For secret-key generation, we use 18 AP-STA pairs and collect 108,000 Wi-Fi packets. During data collection, both CSI and BFI are extracted from every packet. 
Data for the three Wi-Fi-based security applications were collected on different days, with the overall data collection period spanning two months.
This study has received IRB approval. All user and device identities are anonymized, and datasets contain only pseudonymous identifiers. The generated secret keys are used solely for evaluation and not for actual communication.





\textbf{Baselines.}
We compare the performance of BFIAttack against two baseline methods. The first is a random attack, where the adversary generates random CSI values without incorporating any domain knowledge. The second is DomPathCon~\cite{qu2024guessing}, the state-of-the-art attack targeting Wi-Fi-based physical-layer security. DomPathCon is an active attack strategy that guesses the CSI amplitudes of dominant signal paths using a small set of discretized offset values.
We exclude knowledgeable attacks that assume full access to a legitimate user's or device's CSI, as such assumptions are impractical.



\textbf{Evaluation Metrics.}
We evaluate the performance of the attack using the Attack Success Rate (ASR). In each attack trial, the adversary is allowed to perform up to $Q$ attack attempts, each utilizing a different reconstructed CSI derived from the captured BFI. An attack trial is considered successful if the adversary achieves a successful attack within these $Q$ attempts. Formally, the ASR is defined as: $\text{ASR} = \frac{\text{Number of successful attacks within } Q \text{ attempts}}{\text{Total number of attack trials}} \times 100\%$. Unless otherwise specified, the default number of attack attempts is set to $Q=5$.
For device and user authentication, a successful attack is defined as the adversary successfully impersonating the legitimate device or user to the verifier. For secret key generation, a successful attack means that the adversary correctly infers the shared secret key between the devices.

\subsection{Overall Performance}



We first evaluate the overall performance of BFIAttack across three Wi-Fi-based physical-layer security applications (i.e., device authentication, user authentication, and secret-key generation) and compare it against two baselines (i.e., DomPathCon and the Random Attack).
As shown in Table~\ref{tab:asrN_multi}, in the multi-antenna STA scenario, BFIAttack achieves an average ASR of 73\% with no more than 5 attack attempts. Specifically, BFIAttack attains 76.06\% ASR for device authentication, 70.30\% for user authentication, and 72.73\% for secret key generation. In contrast, the average ASR of DomPathCon is only 2.9\%, while the Random Attack achieves a negligible 0.2\%. Furthermore, when the number of attempts is 20, the average ASR of BFIAttack reaches 90\%. 
As shown in Table~\ref{tab:asrN_single}, BFIAttack achieves an average ASR exceeding 93\% with only a single attempt in the single-antenna STA scenario. We note that the number of attempts is only one in this scenario, since the closed-form solution can only reconstruct one CSI.
In contrast, DomPathCon achieves an average ASR of only 4.1\%, while the ASR of Random Attack is zero.

The reason is that DomPathCon performs active guessing in both multi-antenna and single-antenna STA scenarios, while the Random Attack generates CSI randomly. BFIAttack leverages effective MLE-based and closed-form CSI reconstruction methods to conduct attacks. Therefore, BFIAttack achieves a high attack success rate with only a few attempts, significantly outperforming the baseline attacks. Overall, BFIAttack poses a substantially greater threat to Wi-Fi-based physical-layer security applications than existing attack methods.

\begin{table}[t]
\centering
\caption{Attack success rate (\%) in multi-antenna STA scenario compared with two baselines.}
\label{tab:asrN_multi}
\footnotesize
\setlength{\tabcolsep}{4pt}
\renewcommand{\arraystretch}{1.08}
{%
\begin{tabular}{clccc}
\toprule
\textbf{Attempts ($Q$)} & \textbf{Security Application} &
\textbf{BFIAttack} &
\textbf{DomPathCon} &
\textbf{Random Attack} \\
\midrule
\multirow{3}{*}{5}
 & Device Authentication     & \textbf{76.06} & 3.64 & 0.30 \\
 & User Authentication       & \textbf{70.30} & 2.42 & 0.00 \\
 & Secret Key Generation     & \textbf{72.73} & 2.73 & 0.30 \\
\midrule
\multirow{3}{*}{10}
 & Device Authentication     & \textbf{84.55} & 7.58 & 0.91 \\
 & User Authentication       & \textbf{75.15} & 5.45 & 0.30 \\
 & Secret Key Generation     & \textbf{79.39} & 5.15 & 0.61 \\
\midrule
\multirow{3}{*}{15}
 & Device Authentication     & \textbf{88.48} & 10.30 & 2.12 \\
 & User Authentication       & \textbf{82.72} & 6.36 & 0.61 \\
 & Secret Key Generation     & \textbf{83.33} & 6.67 & 0.97 \\
\midrule
\multirow{3}{*}{20}
 & Device Authentication     & \textbf{93.03} & 14.55 & 2.58 \\
 & User Authentication       & \textbf{88.18} & 10.30 & 0.97 \\
 & Secret Key Generation     & \textbf{88.79} & 9.70 & 1.29 \\
\bottomrule
\end{tabular}}
\end{table}

\begin{table}[t]
\centering
\caption{Attack success rate (\%) in single-antenna STA scenario compared with two baselines.}
\label{tab:asrN_single}
\footnotesize
\setlength{\tabcolsep}{4pt}
\renewcommand{\arraystretch}{1.08}
{%
\begin{tabular}{clccc}
\toprule
\textbf{Attempts ($Q$)} & \textbf{Security Application} &
\textbf{BFIAttack} &
\textbf{DomPathCon} &
\textbf{Random Attack} \\
\midrule
\multirow{3}{*}{1}
 & Device Authentication     & \textbf{95.48} & 6.45 & 0.00 \\
 & User Authentication       & \textbf{92.58} & 1.94 & 0.00 \\
 & Secret Key Generation     & \textbf{94.19} & 3.87 & 0.00 \\
\bottomrule
\end{tabular}}
\end{table}

\subsection{Impact of Different Numbers of Antennas}

\begin{figure}[t]
  \centering
  \begin{subfigure}[b]{0.4\linewidth}
    \includegraphics[width=\linewidth]{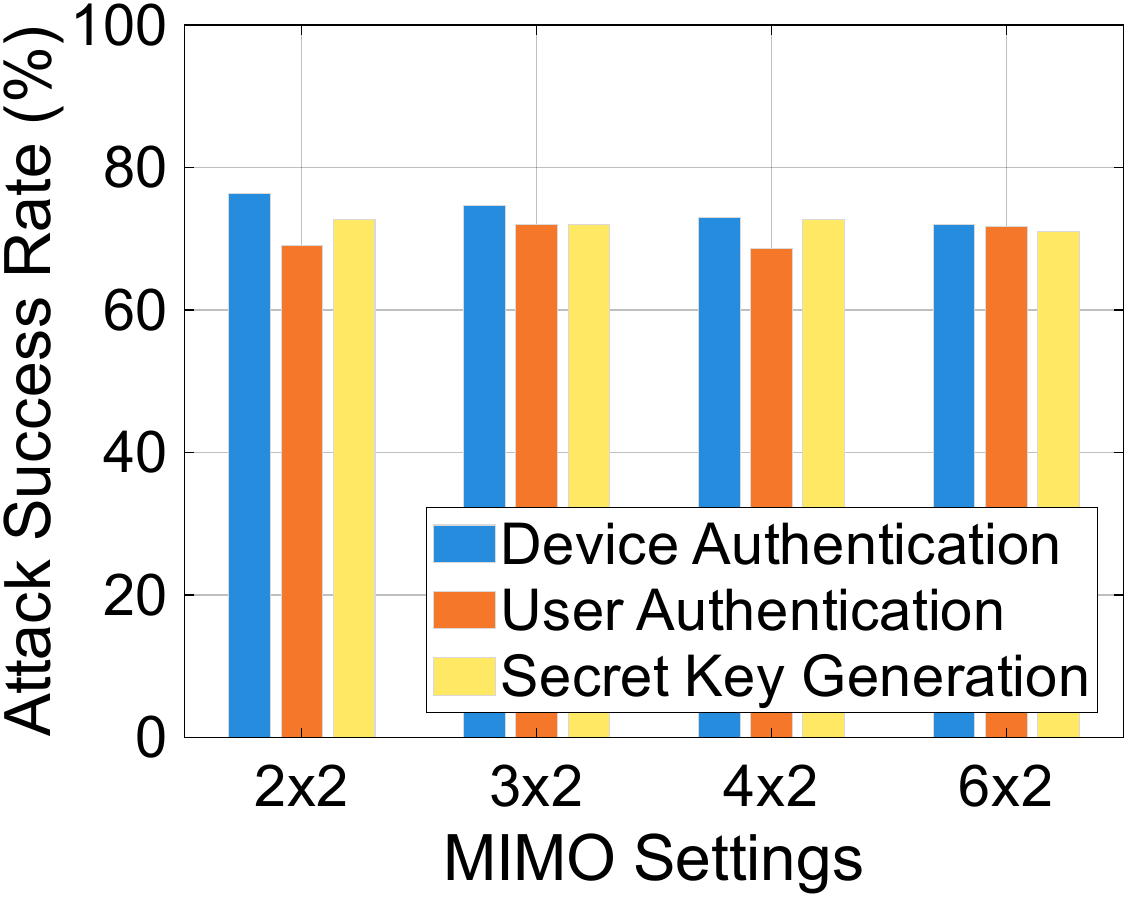}
    \caption{Multi-antenna STA scenario.}
    \label{fig:a}
  \end{subfigure}\hspace{0.05\linewidth}
  \begin{subfigure}[b]{0.4\linewidth}
    \includegraphics[width=\linewidth]{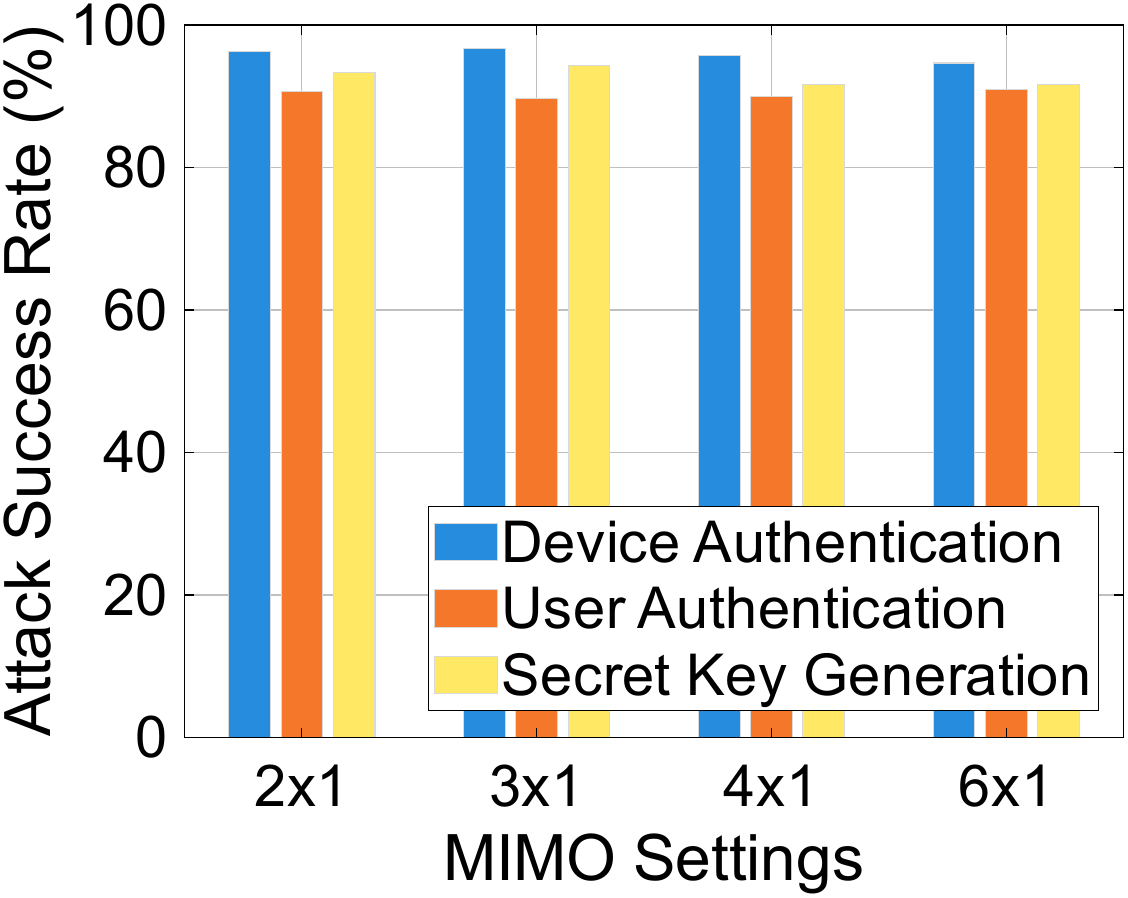}
    \caption{Single-antenna STA scenario.}
    \label{fig:b}
  \end{subfigure}
  \caption{Impact of the number of antennas.}
  \label{fig:dif_antenna}
\end{figure}

To evaluate the impact of different numbers of antennas, we vary the MIMO settings in the multi-antenna STA scenario from $2 \times 2$, $3 \times 2$, $4 \times 2$, to $6 \times 2$, and in the single-antenna STA scenario from $2 \times 1$, $3 \times 1$, $4 \times 1$, to $6 \times 1$. The results are shown in Figure~\ref{fig:dif_antenna}. We observe no significant variation in ASR across different antenna configurations in either scenario. 
Specifically, in the multi-antenna STA scenario, the ASR remains approximately 72\%-76\% for device authentication, 69\%-72\% for user authentication, and 71\%-73\% for secret key generation. In the single-antenna STA scenario, the ASR consistently exceeds 89\% across all three applications, with device authentication around 95\%-97\%, user authentication around 90\%-91\%, and secret key generation around 92\%-94\%. These results indicate that both the MLE-based and closed-form CSI reconstruction methods, as well as our attack strategies, remain effective in both scenarios regardless of antenna configuration. BFIAttack therefore demonstrates strong applicability across a wide range of antenna settings.


\subsection{Impact of Different Bandwidths}
\begin{figure}[t]
  \centering
  \begin{subfigure}[b]{0.4\linewidth}
    \includegraphics[width=\linewidth]{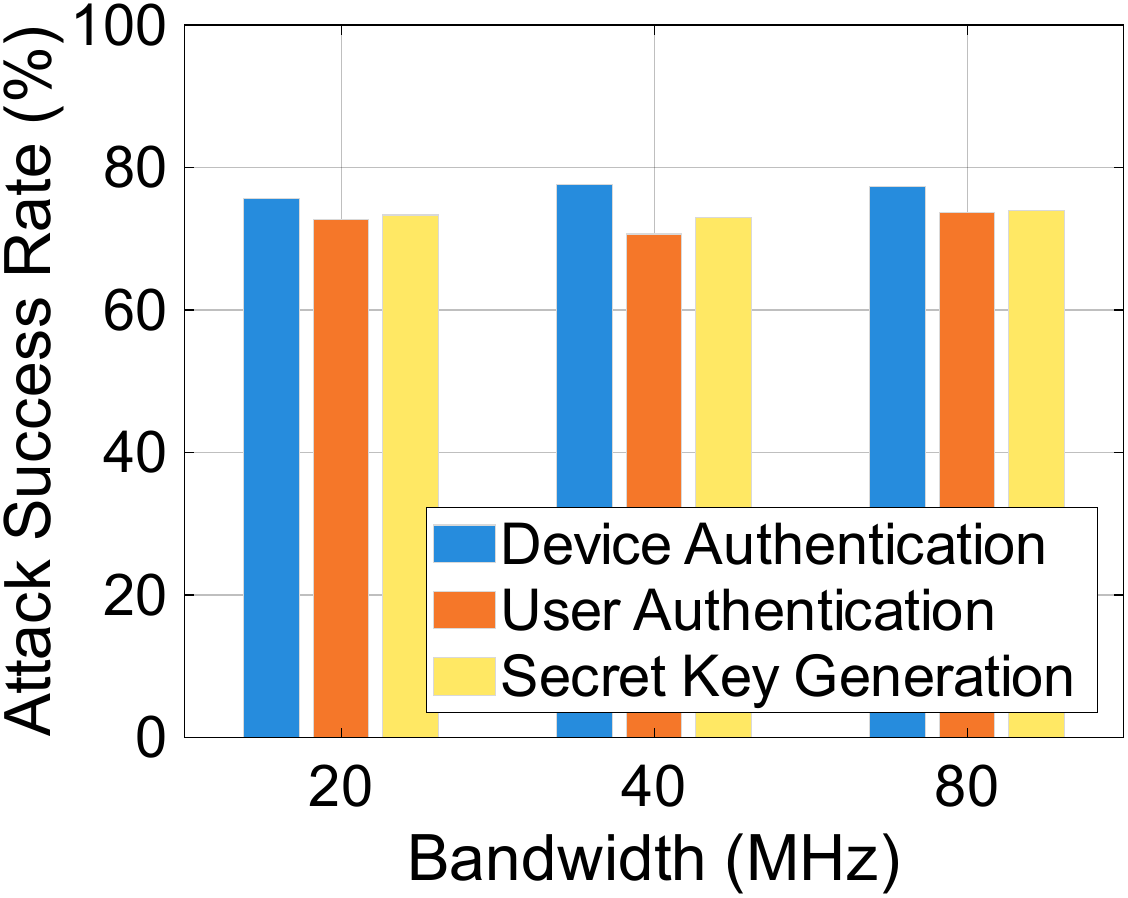}
    \caption{Multi-antenna STA scenario.}
    \label{fig:a}
  \end{subfigure}\hspace{0.05\linewidth}
  \begin{subfigure}[b]{0.4\linewidth}
    \includegraphics[width=\linewidth]{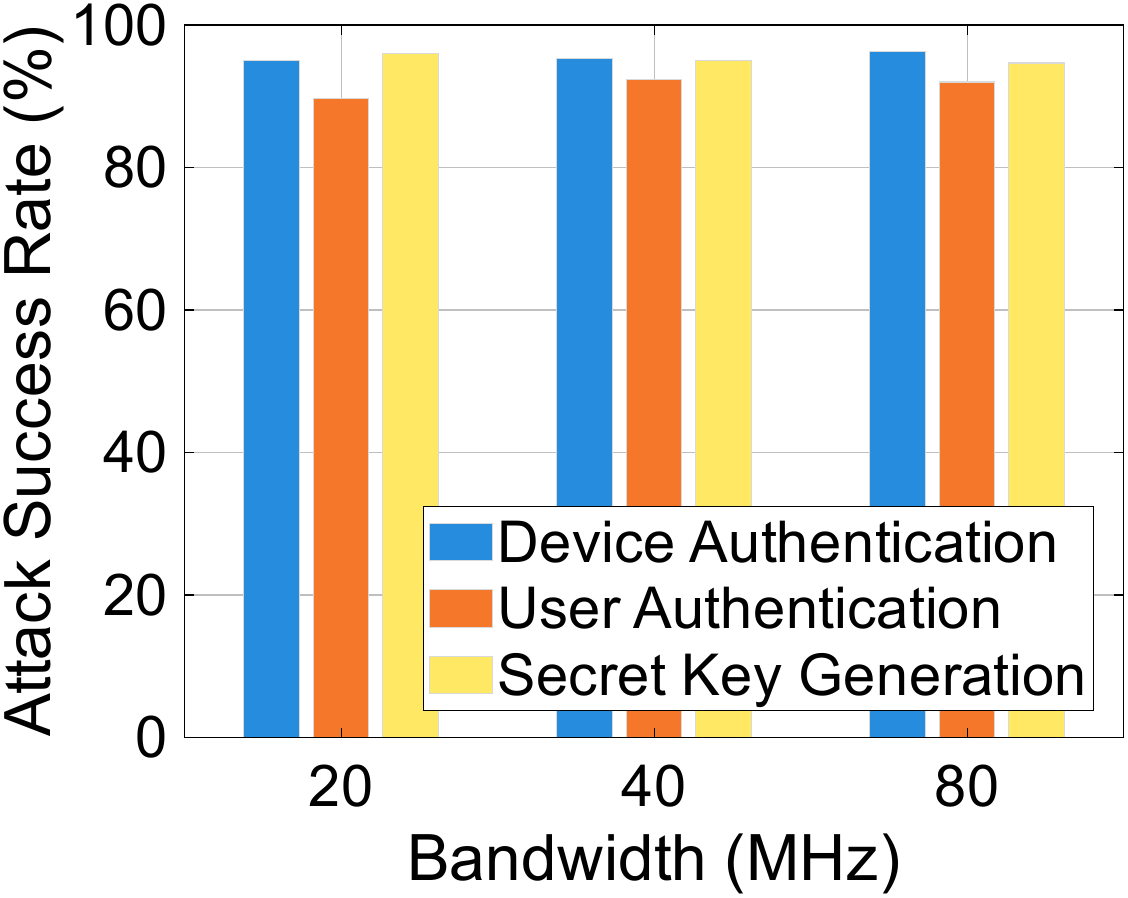}
    \caption{Single-antenna STA scenario.}
    \label{fig:b}
  \end{subfigure}
  \caption{Impact of different bandwidths.}
  \label{fig:dif_ban}
\end{figure}

We also evaluate the impact of bandwidth by varying it among 20 MHz, 40 MHz, and 80 MHz. As shown in Figure~\ref{fig:dif_ban}, there are no significant changes in ASR across different bandwidths in either the multi-antenna or single-antenna STA scenarios. 
Across all three security applications in both scenarios, the ASR varies by less than 3\%, with no consistent increasing or decreasing trend.
This robustness arises because varying the bandwidth changes the number of subcarriers but does not affect BFI extraction, and therefore does not impact the effectiveness of our attack.
This result indicates that BFIAttack is robust to bandwidth variations and can effectively target a wide range of devices operating under different bandwidths.

\subsection{Impact of Different Devices}
\begin{figure}[t]
  \centering
  \begin{subfigure}[b]{0.4\linewidth}
    \includegraphics[width=\linewidth]{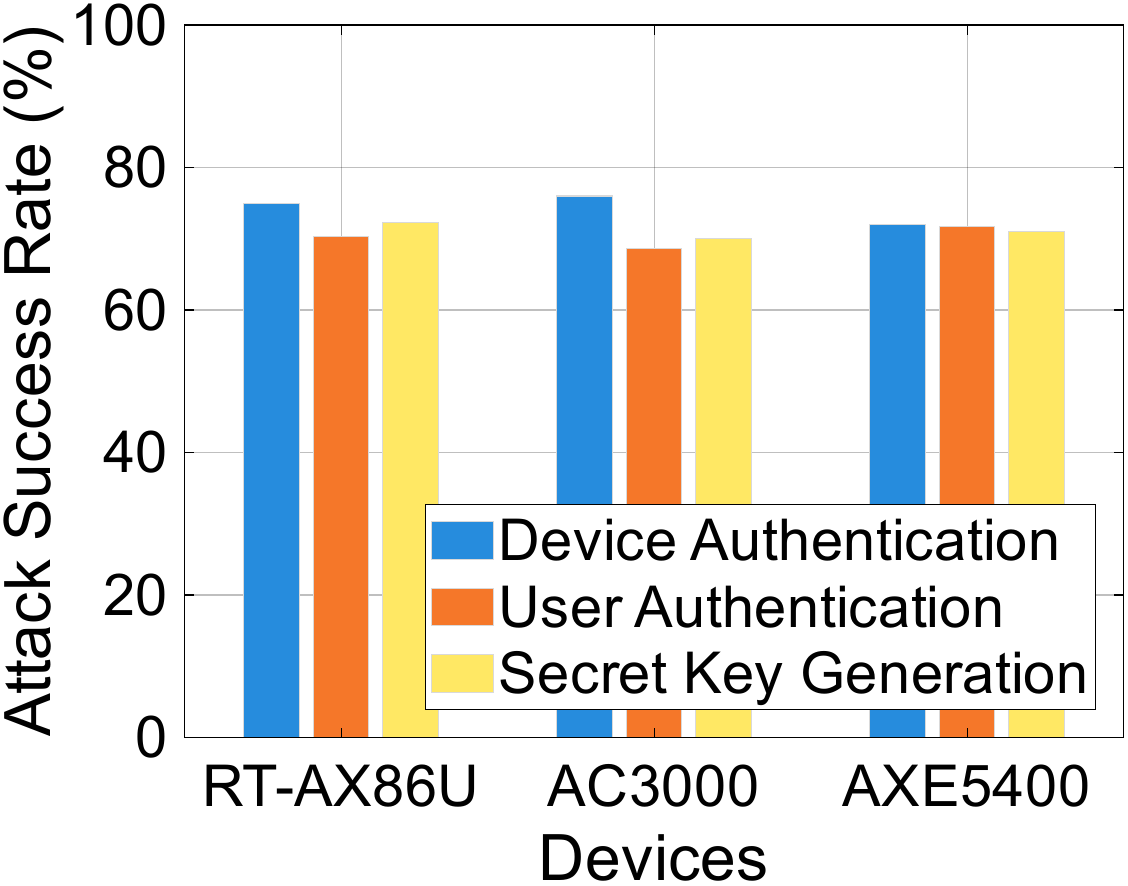}
    \caption{Multi-antenna STA scenario.}
    \label{fig:a}
  \end{subfigure}\hspace{0.05\linewidth}
  \begin{subfigure}[b]{0.4\linewidth}
    \includegraphics[width=\linewidth]{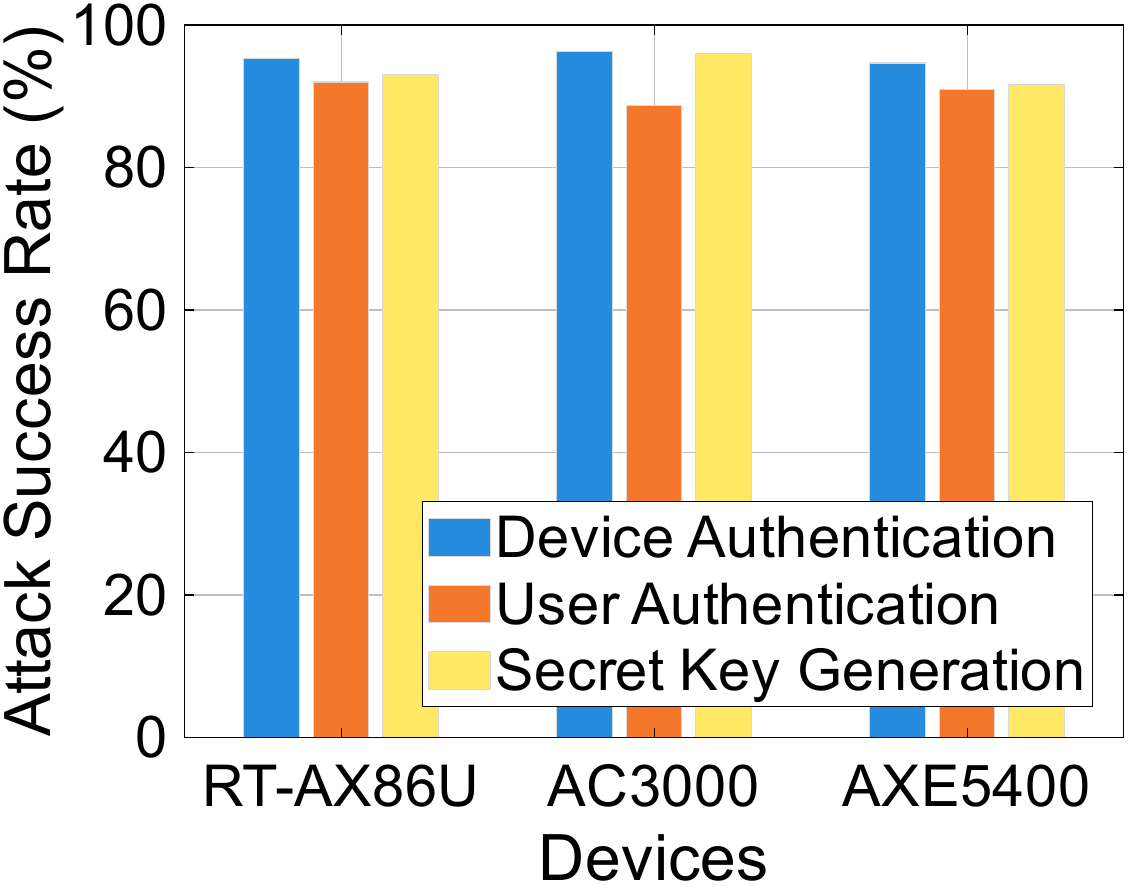}
    \caption{Single-antenna STA scenario.}
    \label{fig:b}
  \end{subfigure}
  \caption{Impact of different devices.}
  \label{fig:dif_devices}
\end{figure}

To evaluate the impact of different Wi-Fi devices, we use three Wi-Fi routers: ASUS RT-AX86U, Linksys AC3000, and TP-Link AXE5400. The results, shown in Figure~\ref{fig:dif_devices}, indicate that the ASR remains consistent across all devices. Specifically, in the multi-antenna STA scenario, ASRs for all three Wi-Fi-based security applications are approximately 72\%-76\%, while in the single-antenna STA scenario, ASRs are approximately 90\%-96\%. These results suggest that differences in device implementations have only a minor impact on the effectiveness of BFIAttack. This is because such differences mainly arise from hardware configurations and chipset designs, which do not affect BFI as long as devices comply with the IEEE 802.11ac/ax standards. Overall, the results demonstrate that the proposed BFIAttack is device-agnostic.

\subsection{Impact of Different Distances}



\begin{figure}[t]
  \centering
  \begin{subfigure}[b]{0.4\linewidth}
    \includegraphics[width=\linewidth]{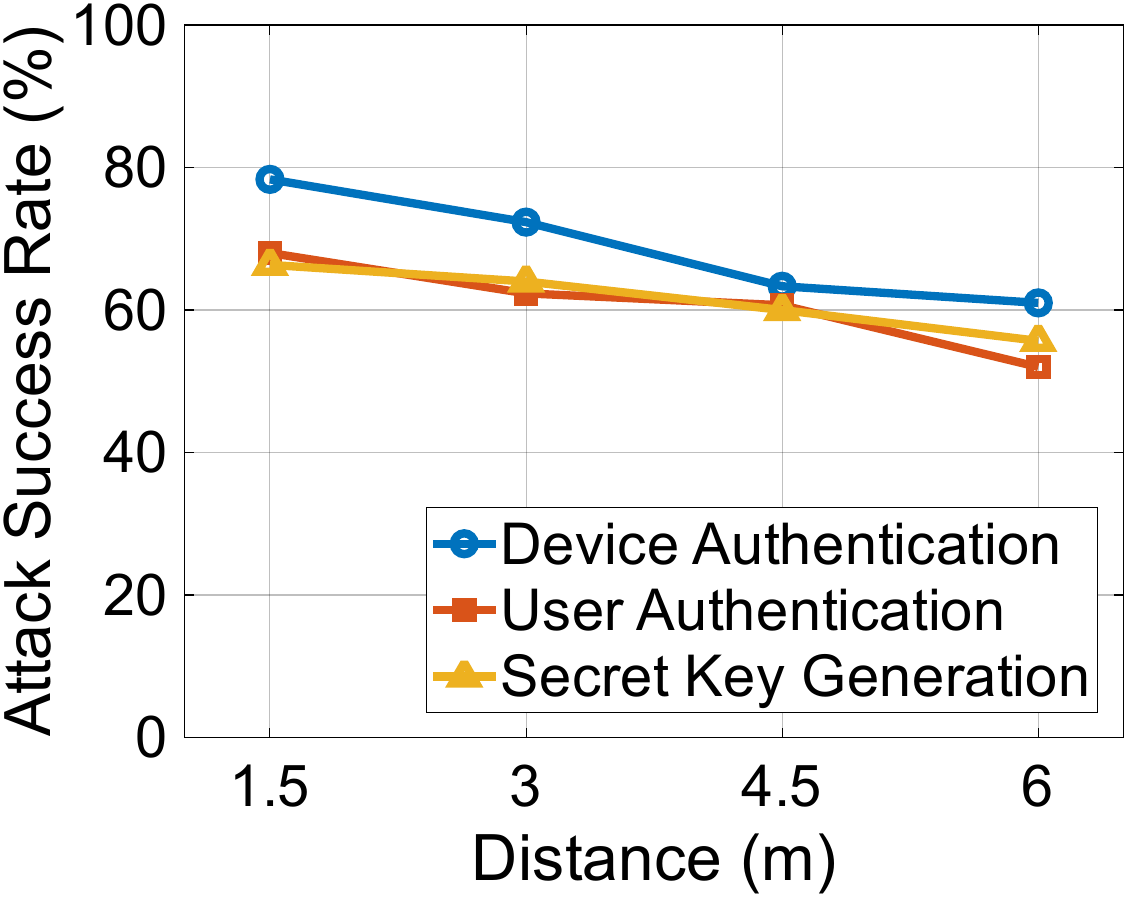}
    \caption{\mbox{Multi-antenna} STA scenario.}
    \label{fig:a}
  \end{subfigure}\hspace{0.05\linewidth}
  \begin{subfigure}[b]{0.4\linewidth}
    \includegraphics[width=\linewidth]{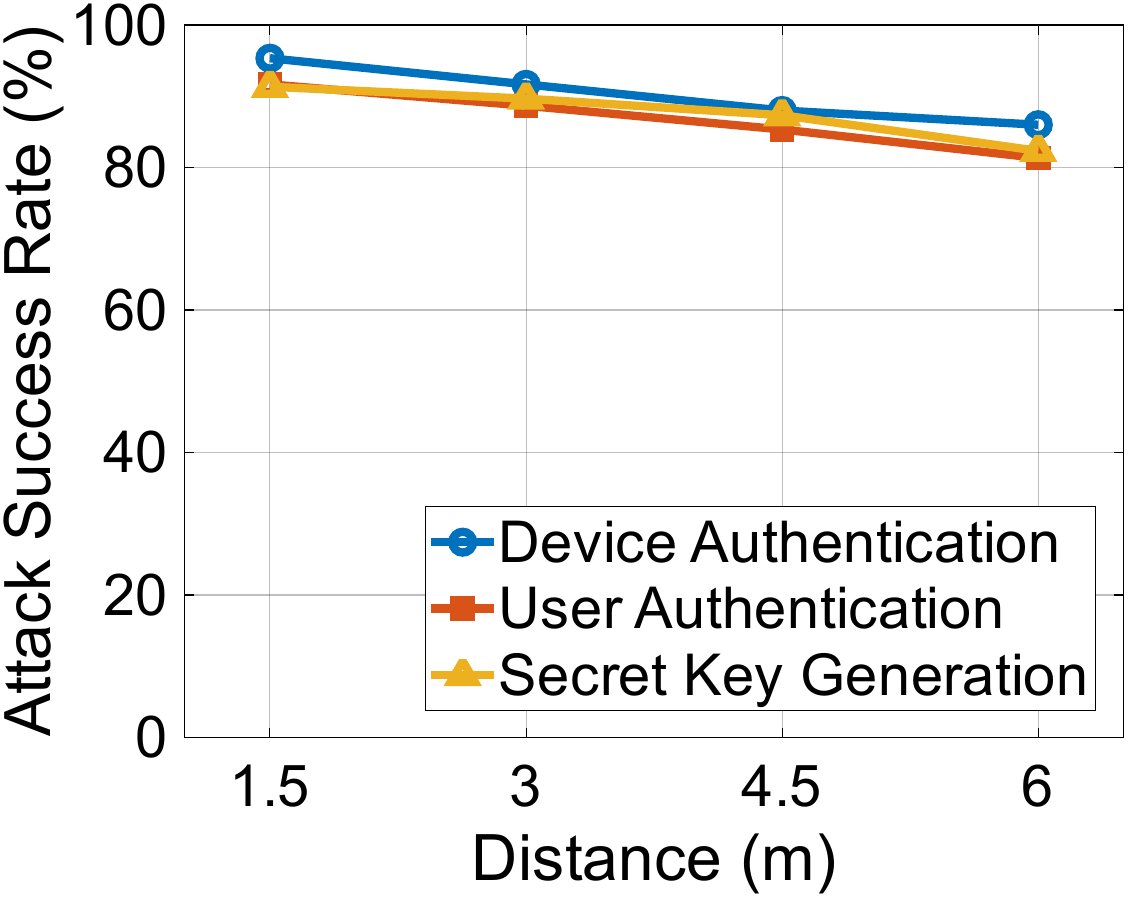}
    \caption{\mbox{Single-antenna} STA scenario.}
    \label{fig:b}
  \end{subfigure}
  \caption{Impact of different distances.}
  \label{fig:dif_dis}
\end{figure}

We evaluate the performance of BFIAttack under different distances between the sniffer and the AP: 1.5\,m, 3\,m, 4.5\,m, and 6\,m. As shown in Figure~\ref{fig:dif_dis}, decreasing the distance from 6\,m to 1.5\,m leads to an increase in ASR for both scenarios. 
In the single-antenna STA case, ASR improves from approximately 80\% to 95\%, while in the multi-antenna STA case, it increases from about 60\% to 80\%. 
Specifically, as the sniffer moves closer to the AP, ASR for device authentication increases from 61\% to 78\% in the multi-antenna STA scenario and from 86\% to 95\% in the single-antenna STA scenario. ASR for user authentication rises from approximately 52\% to 68\% and from 81\% to 92\% in the two scenarios, respectively, while ASR for secret key generation increases from about 56\% to 66\% and from 82\% to 91\%. 
The reason is that a shorter sniffer-to-AP distance enables more reliable BFI capture. Nevertheless, our attack remains effective within typical room-scale environments.




\subsection{Impact of Different Environments}

\begin{figure}[t]
  \centering
  \begin{subfigure}[b]{0.4\linewidth}
    \includegraphics[width=\linewidth]{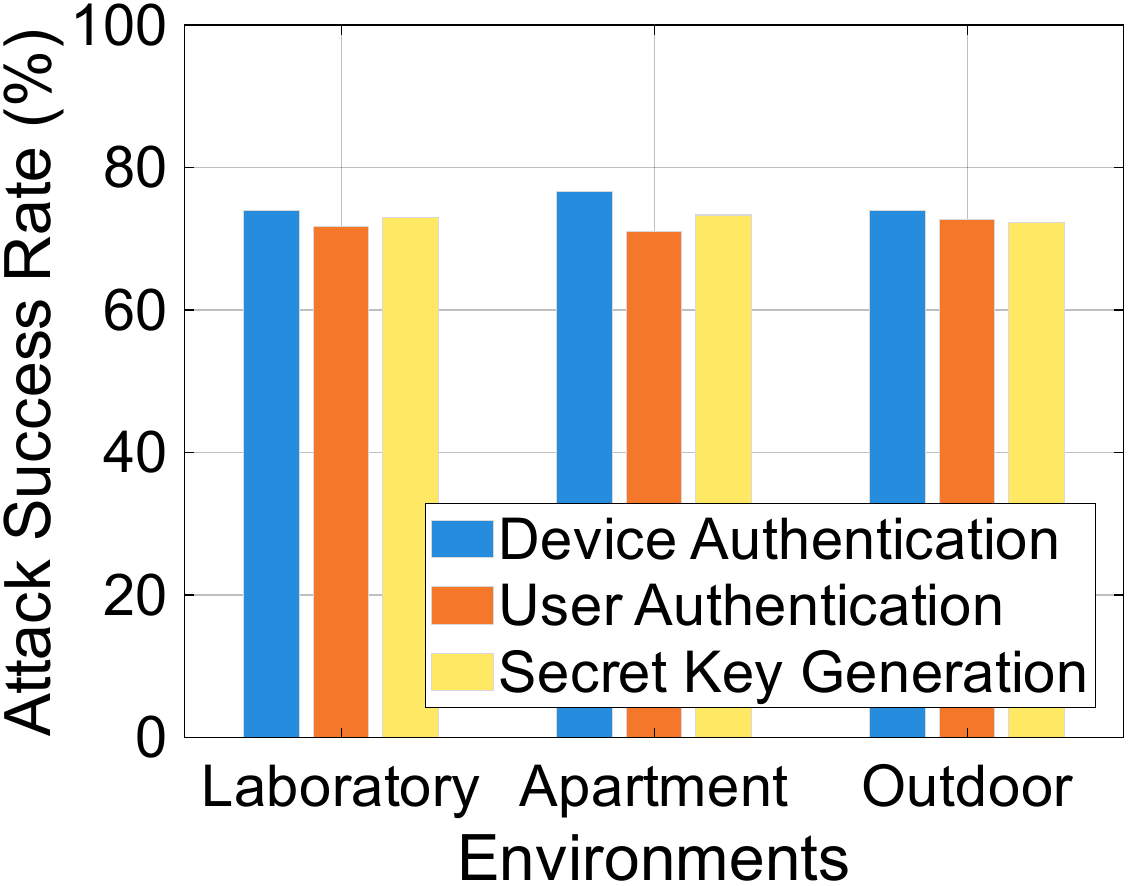}
    \caption{\mbox{Multi-antenna} STA scenario.}
    \label{fig:a}
  \end{subfigure}\hspace{0.05\linewidth}
  \begin{subfigure}[b]{0.4\linewidth}
    \includegraphics[width=\linewidth]{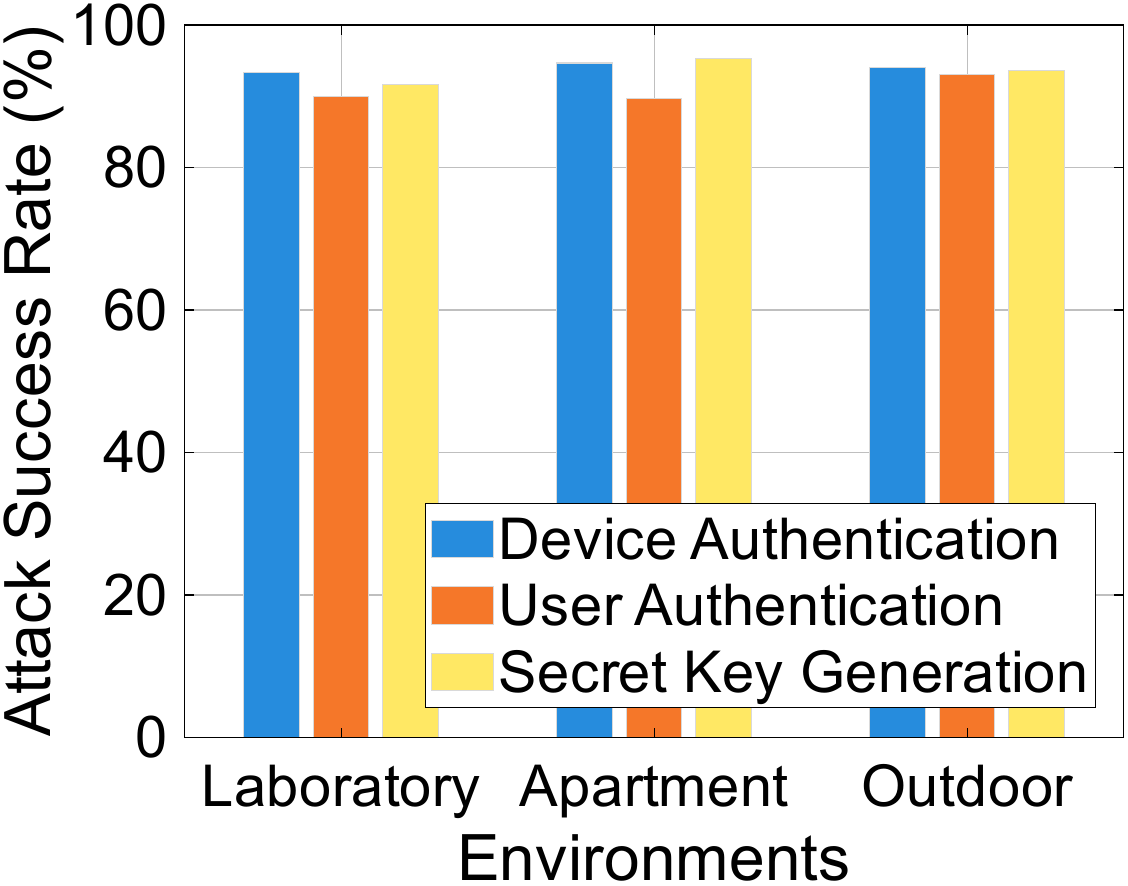}
    \caption{\mbox{Single-antenna} STA scenario.}
    \label{fig:b}
  \end{subfigure}
  \caption{Impact of different environments.}
  \label{fig:dif_env}
\end{figure}

We further evaluate attack performance across diverse environments, including a laboratory, an apartment, and an outdoor area, as shown in Figure~\ref{fig:dif_env}. The results indicate that BFIAttack consistently achieves ASRs of approximately 93\% in the single-antenna STA scenario and 73\% in the multi-antenna STA scenario across all environments. Specifically, in the multi-antenna STA scenario, ASRs for all three applications remain within about 72\%-77\% across the three environments, while in the single-antenna STA scenario, ASRs consistently range from about 90\%-95\% for these three environments. These findings demonstrate that BFIAttack is robust to environmental variations.

\subsection{Impact of Moving People}
\begin{figure}[t]
  \centering
  \begin{subfigure}[b]{0.4\linewidth}
    \includegraphics[width=\linewidth]{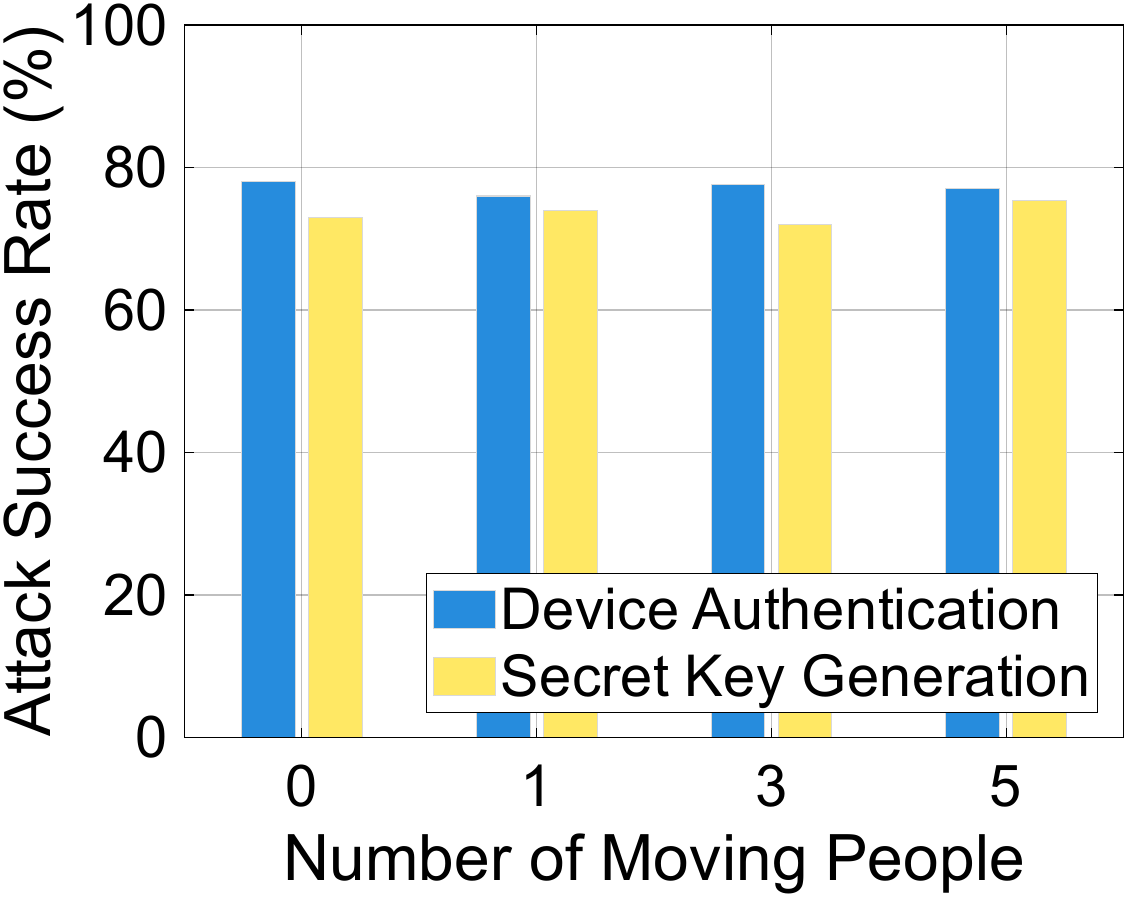}
    \caption{\mbox{Multi-antenna} STA scenario.}
    \label{fig:a}
  \end{subfigure}\hspace{0.05\linewidth}
  \begin{subfigure}[b]{0.4\linewidth}
    \includegraphics[width=\linewidth]{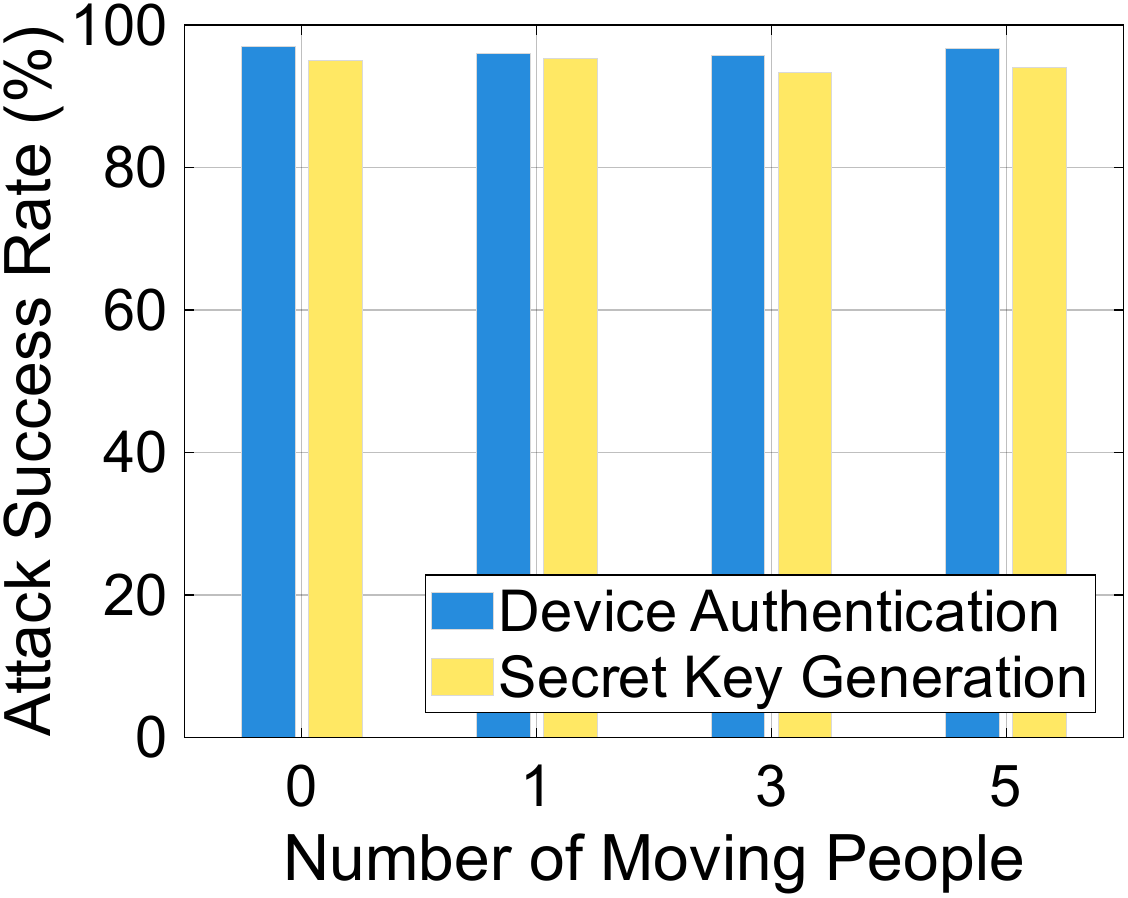}
    \caption{\mbox{Single-antenna} STA scenario.}
    \label{fig:b}
  \end{subfigure}
  \caption{Impact of moving people.}
  \label{fig:dif_user}
\end{figure}

In this evaluation, we investigate how the presence of moving people affects the attack performance. Specifically, we ask 1, 3, and 5 people to move randomly within the environment while the legitimate devices perform security applications (i.e., device authentication and secret key generation), and the adversary passively sniffs the BFI simultaneously. We exclude the evaluation of user authentication under these conditions, as the user's gait-related CSI can be significantly distorted by surrounding human motion, rendering Wi-Fi-based gait authentication infeasible.
As shown in Figure~\ref{fig:dif_user}, in the multi-antenna STA scenario, the ASRs for device authentication and secret key generation remain at approximately 77\% and 75\%, respectively, even when five people are moving. Similarly, in the single-antenna STA scenario, the attack success rates for device authentication and secret key generation remain high at approximately 97\% and 94\%, respectively, even when five people are moving simultaneously. These results demonstrate that the presence of moving individuals has minimal impact on attack performance. This robustness arises because BFI inherently captures channel variations, enabling accurate CSI reconstruction and effective attacks even in dynamic environments.



\subsection{Impact of Non-Line of Sight}

\begin{figure}[t]
  \centering
  \begin{subfigure}[b]{0.4\linewidth}
    \includegraphics[width=\linewidth]{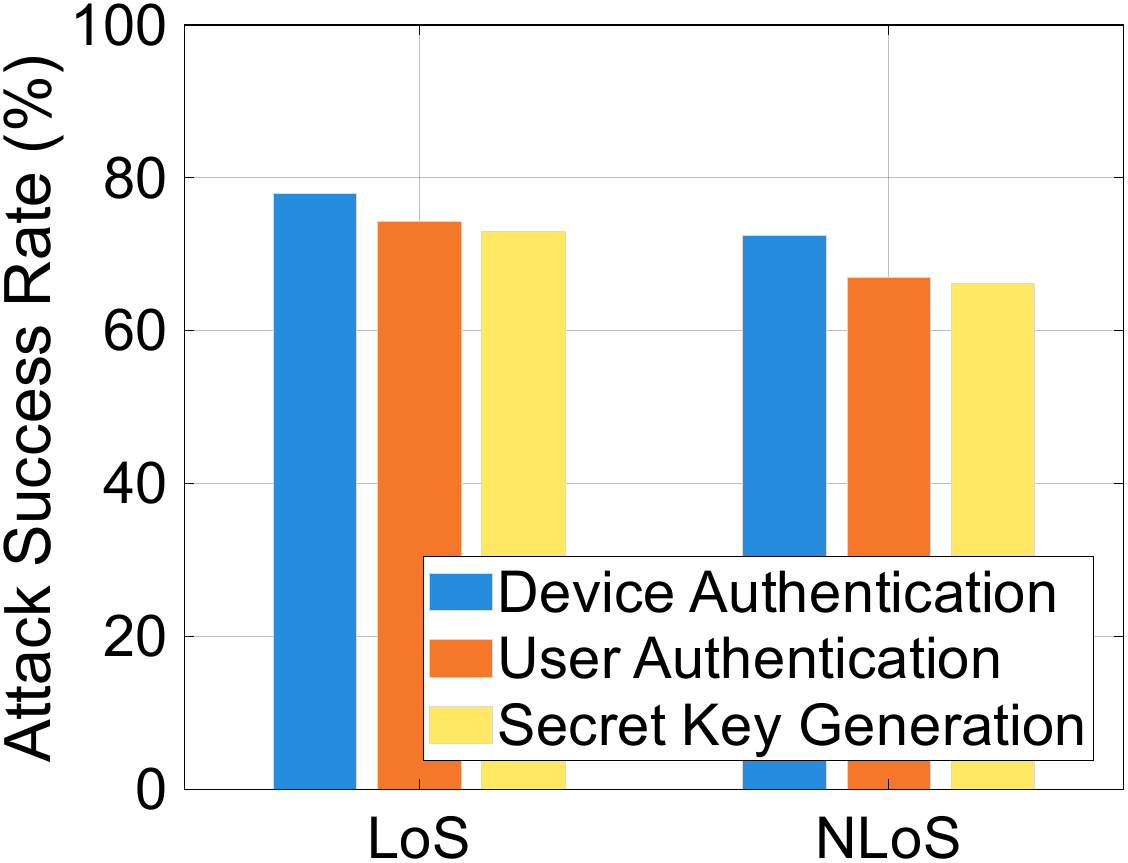}
    \caption{\mbox{Multi-antenna} STA scenario.}
    \label{fig:a}
  \end{subfigure}\hspace{0.05\linewidth}
  \begin{subfigure}[b]{0.4\linewidth}
    \includegraphics[width=\linewidth]{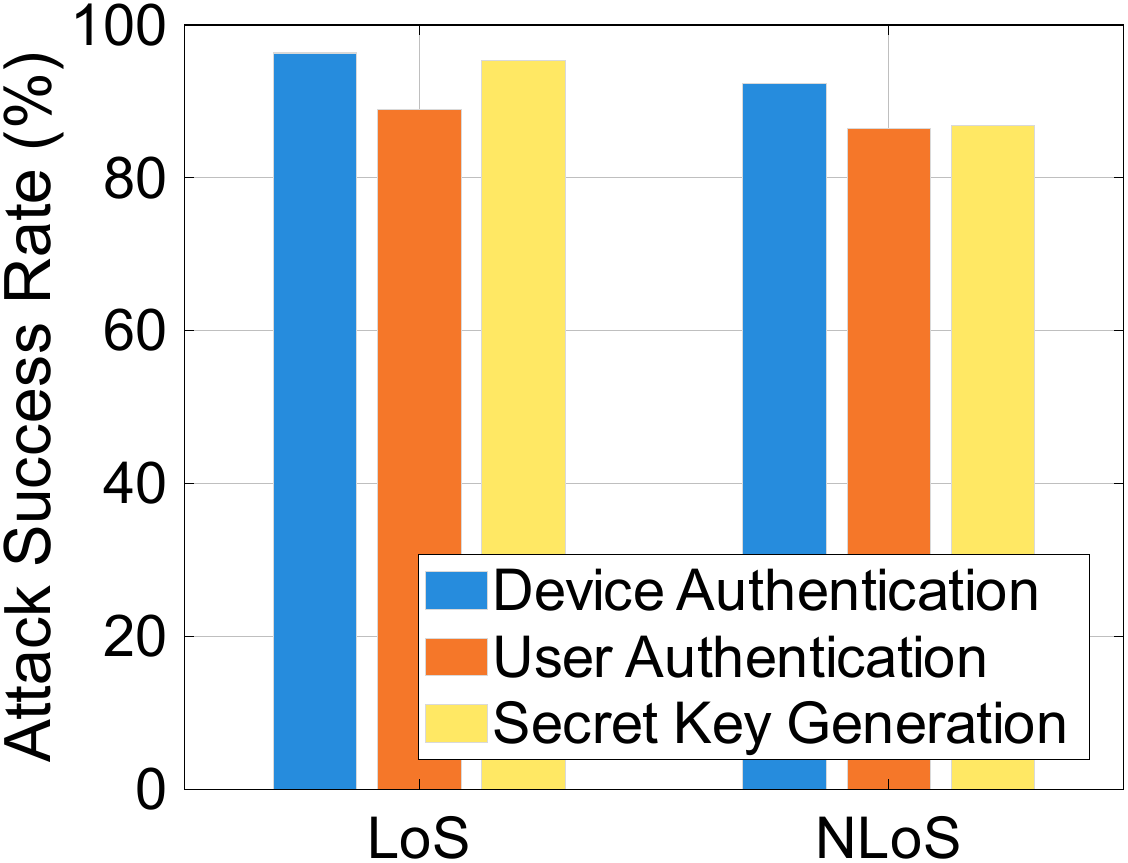}
    \caption{\mbox{Single-antenna} STA scenario.}
    \label{fig:b}
  \end{subfigure}
  \caption{Impact of NLoS conditions.}
  \label{fig:dif_ob}
\end{figure}

To evaluate the impact of non-line-of-sight (NLoS) conditions on the performance of BFIAttack, we conduct experiments in which the sniffer is separated from the AP-STA pair by a wall. As shown in Figure~\ref{fig:dif_ob}, the ASR under the line-of-sight (LoS) condition is slightly higher than that in the NLoS condition. 
Specifically, in the multi-antenna STA scenario, the ASRs for device authentication, user authentication, and secret key generation are approximately 78\%, 74\%, and 73\%, respectively, under LoS conditions, and about 73\%, 67\%, and 66\%, respectively, under NLoS conditions. Similarly, in the single-antenna STA scenario, the corresponding ASRs are around 96\%, 89\%, and 95\% under LoS conditions, and approximately 92\%, 87\%, and 87\% under NLoS conditions.
This slight degradation is likely because the wall partially blocks the BFI transmission.
Nevertheless, these results show that BFIAttack remains effective even in NLoS conditions.

\subsection{Impact of Different Packet Rates}



\begin{figure}[t]
  \centering
  \begin{subfigure}[b]{0.4\linewidth}
    \includegraphics[width=\linewidth]{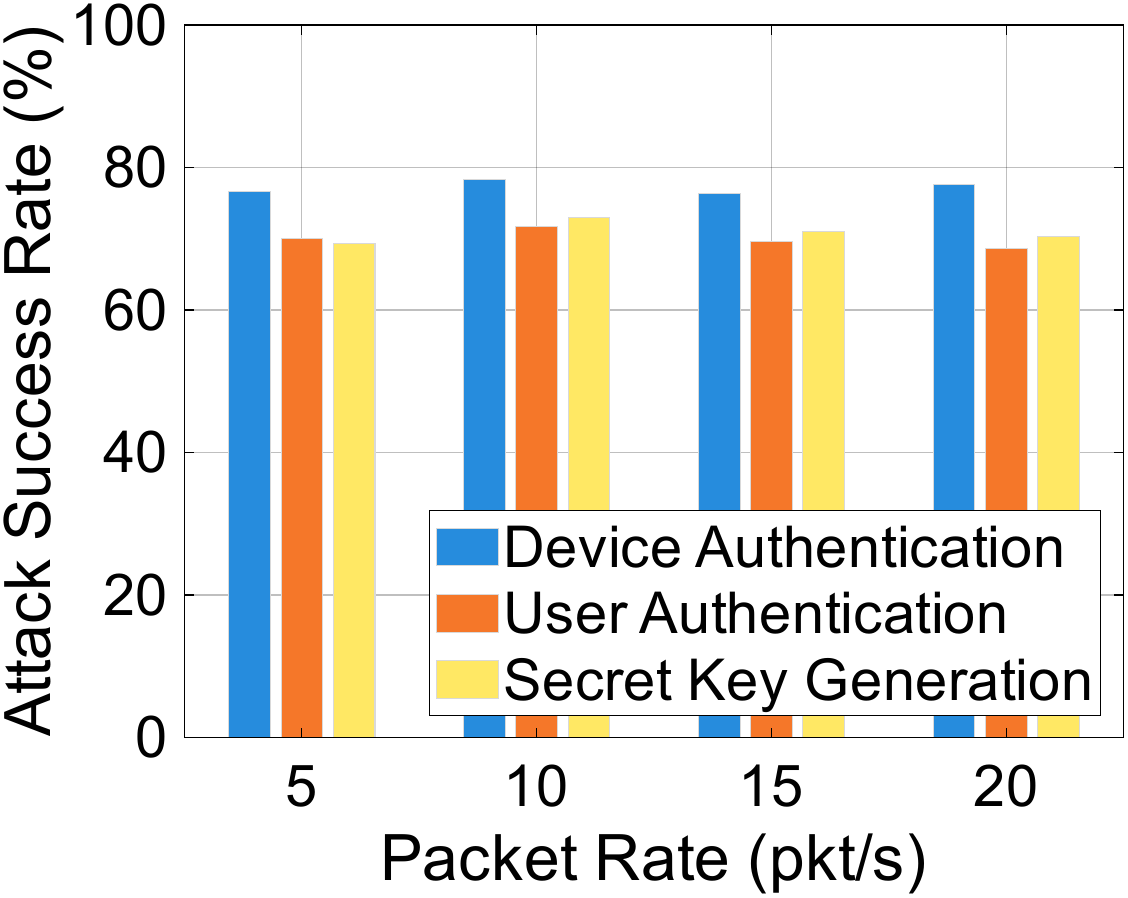}
    \caption{\mbox{Multi-antenna} STA scenario.}
    \label{fig:a}
  \end{subfigure}\hspace{0.05\linewidth}
  \begin{subfigure}[b]{0.4\linewidth}
    \includegraphics[width=\linewidth]{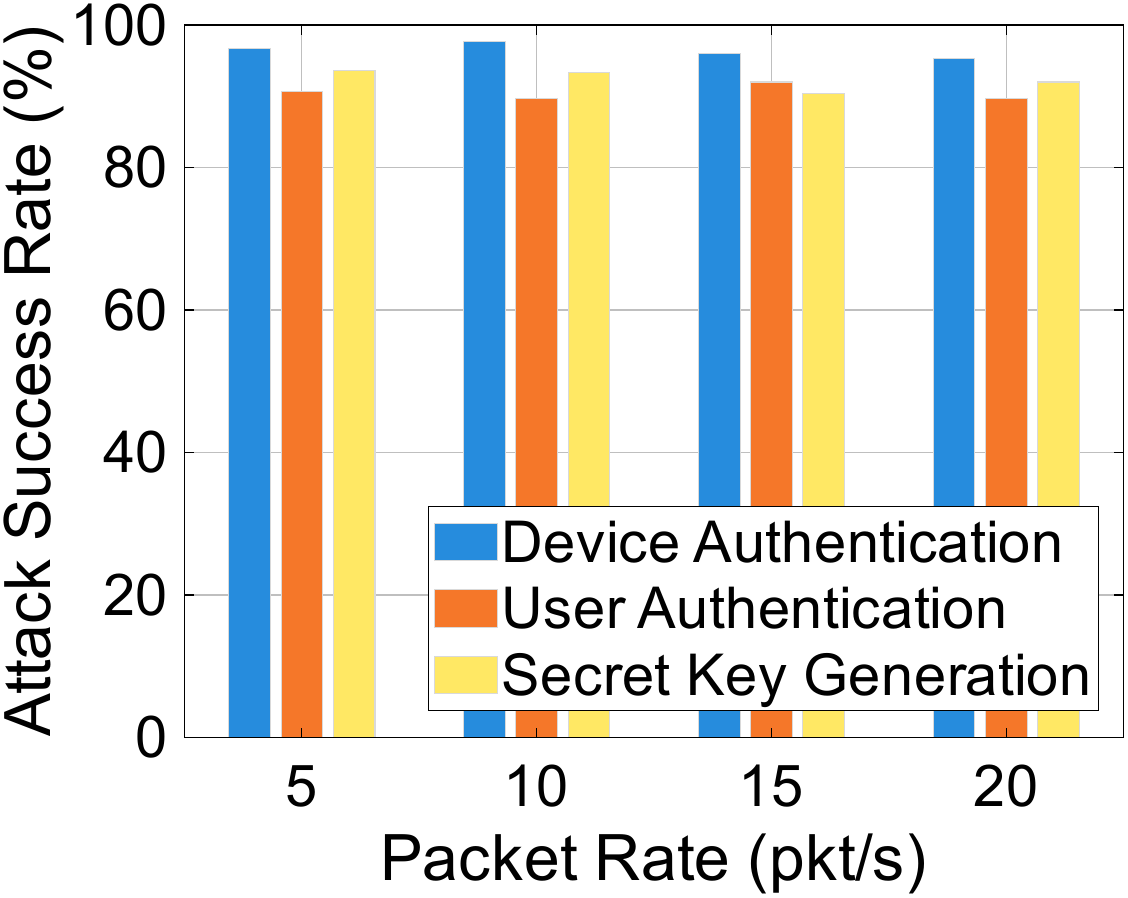}
    \caption{\mbox{Single-antenna} STA scenario.}
    \label{fig:b}
  \end{subfigure}
  \caption{Impact of different packet rates.}
  \label{fig:dif_rat}
\end{figure}

To evaluate the impact of packet rate, we vary the packet rate across 5\,pkt/s, 10\,pkt/s, 15\,pkt/s, and 20\,pkt/s. As shown in Figure~\ref{fig:dif_rat}, the ASR remains almost unchanged across all rates. 
Specifically, the average ASR remains around 93\% in the single-antenna STA scenario and approximately 73\% in the multi-antenna STA scenario. In the single-antenna case, the ASR for device authentication ranges from 95\% to 98\%, the ASR for user authentication remains around 90\%-92\%, and the ASR for secret key generation stays within roughly 90\%-94\%. In the multi-antenna scenario, the corresponding ASRs range from about 76\%-78\%, 69\%-72\%, and 69\%-73\%, respectively.
This invariance is attributed to the fact that varying the packet rate does not affect the capture of BFI. This indicates that BFIAttack's performance is robust to changes in packet rate.



\subsection{Impact of Different Models}
\begin{figure}[t]
  \centering
  \begin{subfigure}[b]{0.4\linewidth}
    \includegraphics[width=\linewidth]{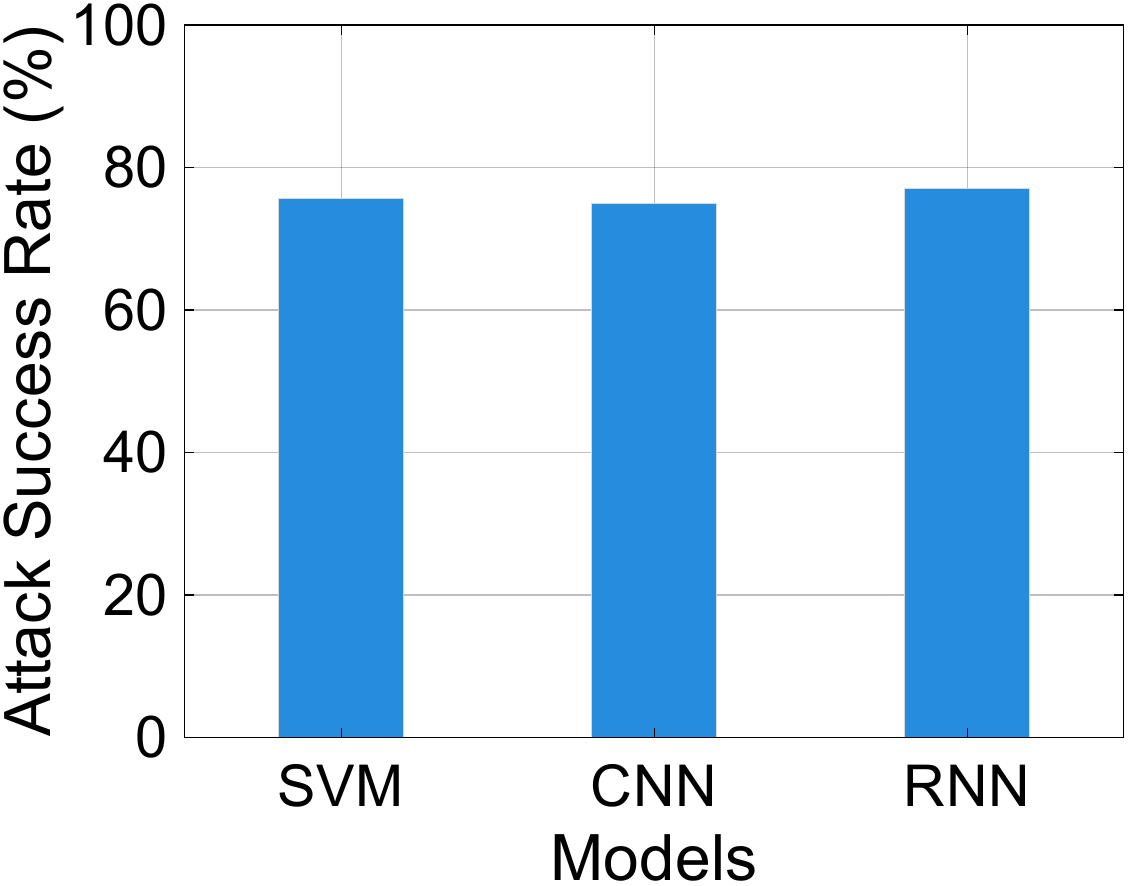}
    \caption{\mbox{Multi-antenna} STA scenario.}
    \label{fig:a}
  \end{subfigure}\hspace{0.05\linewidth}
  \begin{subfigure}[b]{0.4\linewidth}
    \includegraphics[width=\linewidth]{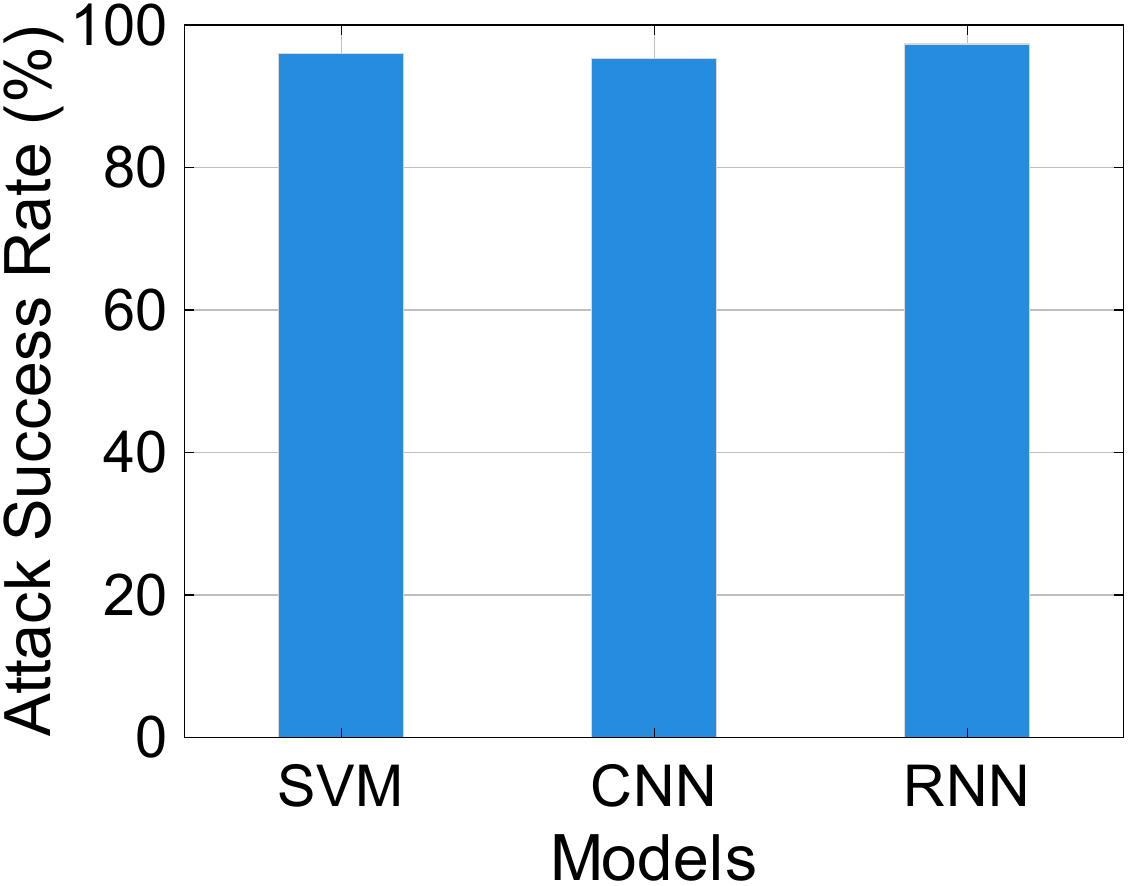}
    \caption{\mbox{Single-antenna} STA scenario.}
    \label{fig:b}
  \end{subfigure}
  \caption{Impact of different models.}
  \label{fig:dif_model}
\end{figure}

We further evaluate the impact of different machine learning models on BFIAttack using device authentication as a representative application. Specifically, we implement SVM, CNN, and RNN models for authentication. As shown in Figure~\ref{fig:dif_model}, we can observe that the ASRs for SVM, CNN, and RNN are 76\%, 75\%, and 77\%, respectively, for multi-antenna STA scenarios. For single-antenna STA scenarios, the ASRs are 96\%, 95\%, and 97\%, respectively.
The ASR remains consistently high across all models in both single-antenna and multi-antenna STA scenarios. It is because BFIAttack reconstructs the fundamental data (i.e., CSI), which is independent of the specific learning model used. These results demonstrate that BFIAttack is robust against different machine learning models.




\subsection{Impact of Time}
\begin{figure}[t]
  \centering

  \begin{minipage}[t]{0.64\linewidth}
    \centering
    \begin{subfigure}[b]{0.45\linewidth}
      \centering
      \includegraphics[width=\linewidth]{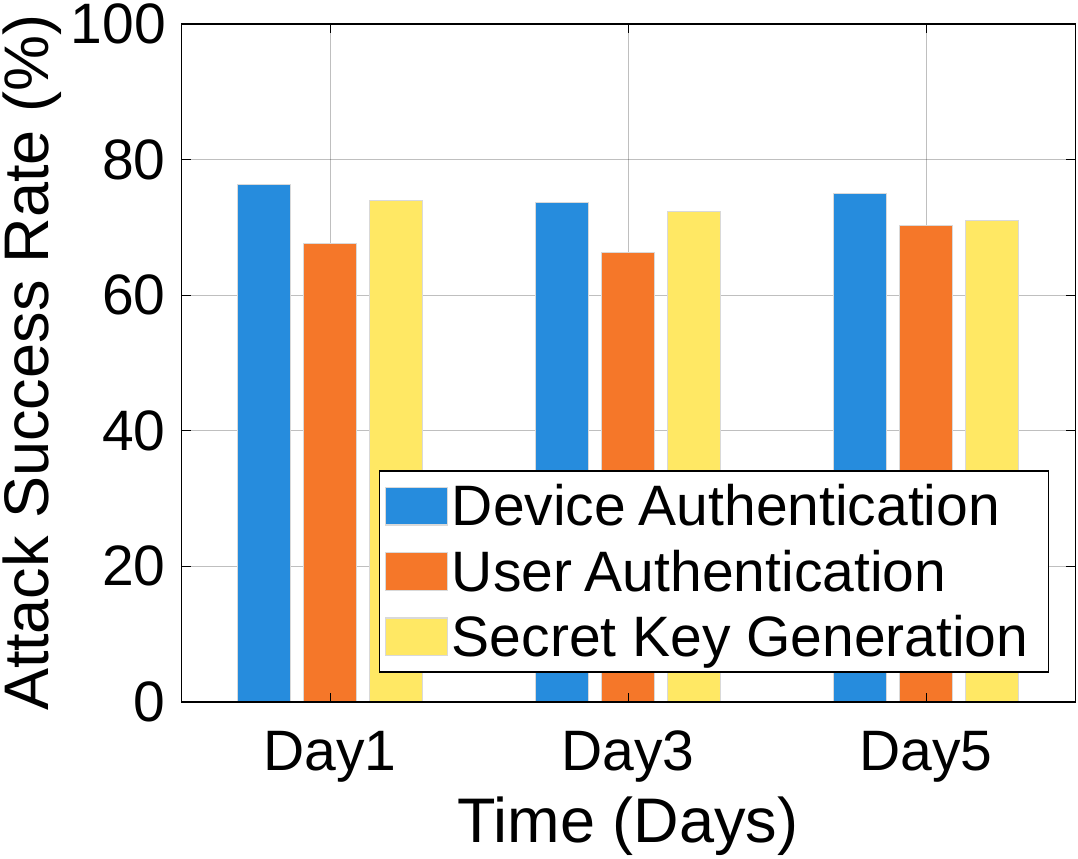}
      \caption{\mbox{Multi-antenna} STA scenario.}
      \label{fig:a}
    \end{subfigure}
    \hspace{0.05\linewidth}
    \begin{subfigure}[b]{0.45\linewidth}
      \centering
      \includegraphics[width=\linewidth]{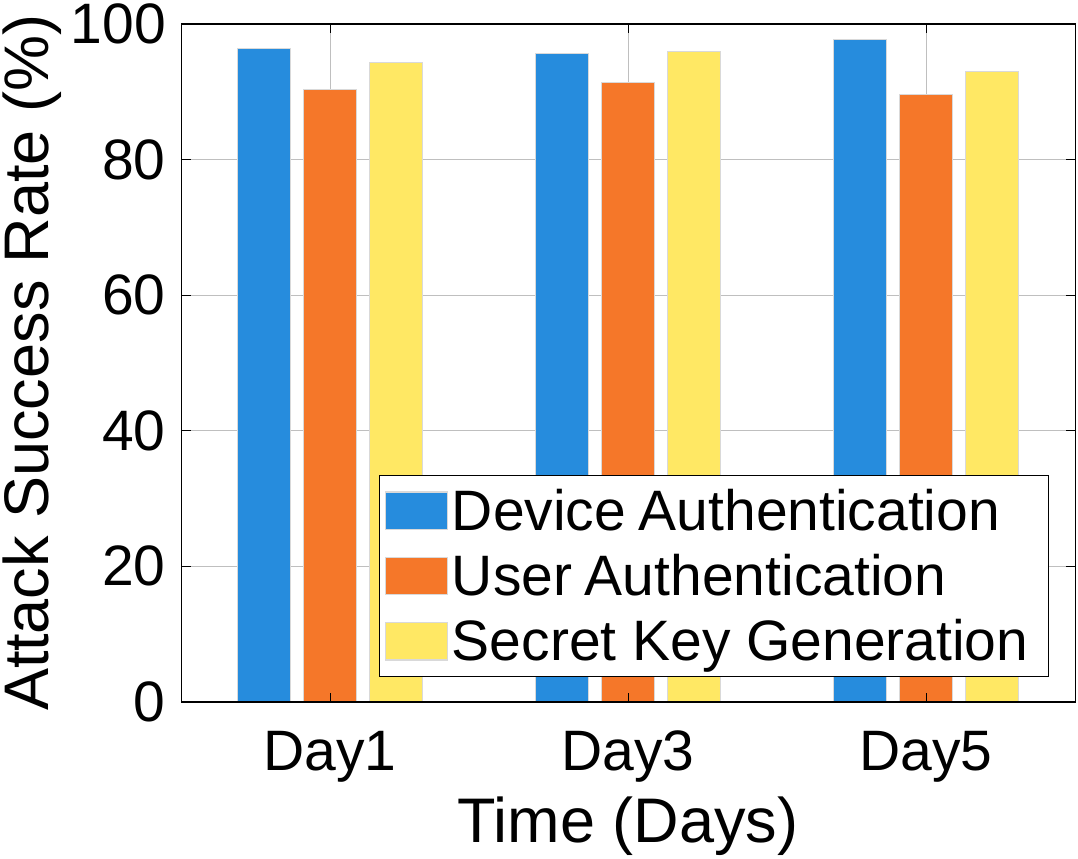}
      \caption{\mbox{Single-antenna} STA scenario.}
      \label{fig:b}
    \end{subfigure}

    \caption{Impact of time.}
    \label{fig:dif_day}
  \end{minipage}
  \hfill
  \begin{minipage}[t]{0.35\linewidth}
    \centering
    \includegraphics[width=\linewidth]{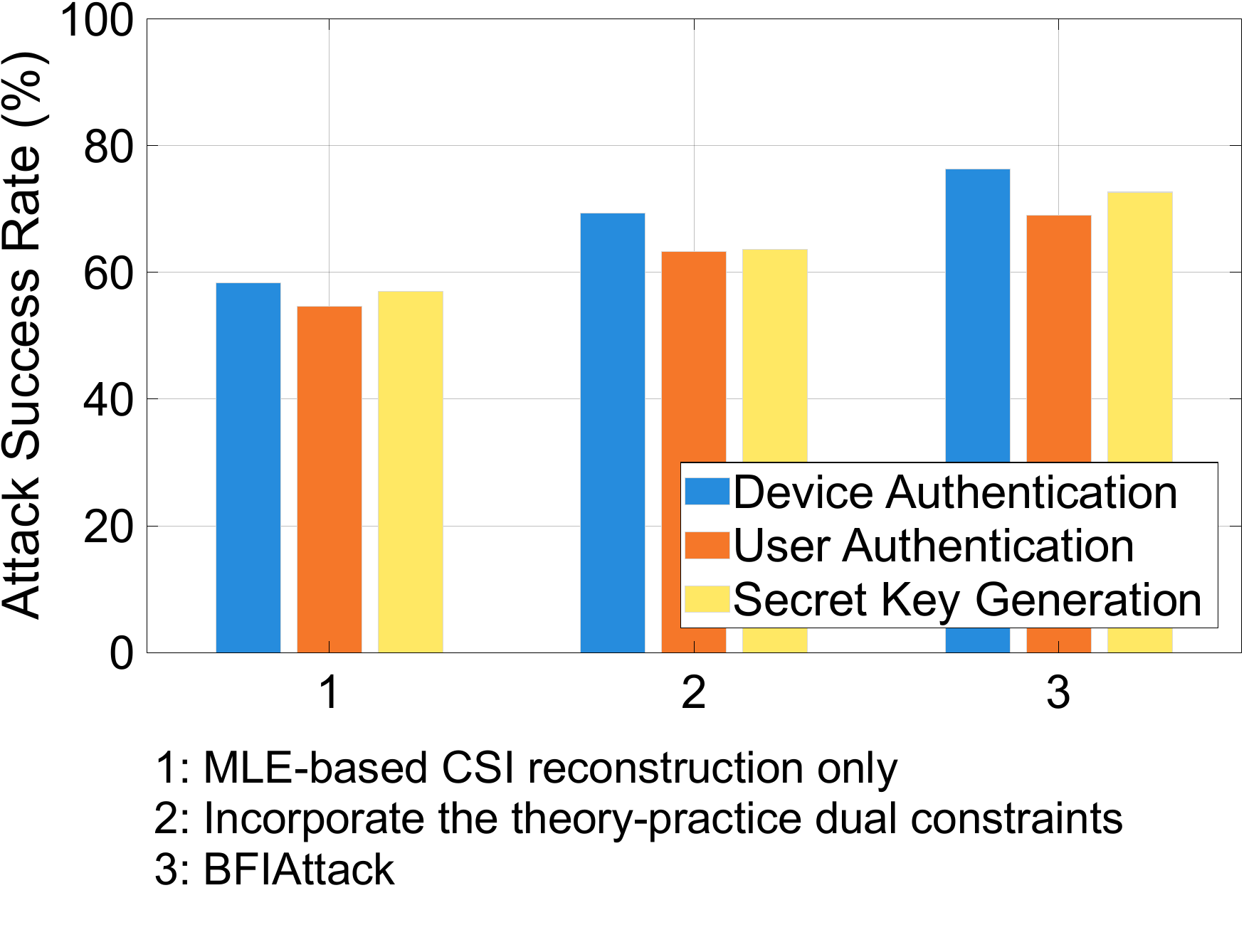}
    \caption{Ablation study.}
    \label{fig:strategy}
  \end{minipage}

\end{figure}

To examine the impact of time on attack performance, we use BFI captured on Day 1 to attack Wi-Fi-based security applications built on CSI collected on Day 1, Day 3, and Day 5. For device authentication and user authentication, the environment (e.g., the locations of the AP and STA) is kept unchanged. For secret key generation, we assume no key updates occur during this period. As shown in Fig.~\ref{fig:dif_day}, for multi-antenna STA scenarios, the average ASRs for Day 1, Day 3, and Day 5 are 74\%, 72\%, and 73\%. For single-antenna STA scenarios, the average ASRs are 94\%, 94\%, and 93\%, respectively. We can observe that our attack exhibits consistent performance over time. This stability arises from an unchanged environment and stable multipath conditions. The evaluation indicates that BFIAttack remains effective over time as long as the environment remains stable.

\subsection{Ablation Study}
To evaluate the contribution of each component in BFIAttack, we conduct an ablation study by adding or removing individual components while keeping the others fixed. This evaluation is performed under the multi-antenna STA scenario. As illustrated in Figure~\ref{fig:strategy}, when only the MLE-based CSI reconstruction is used, the average ASR is 57\%. Incorporating the theory-practice dual constraints increases the ASR to 65\%. Finally, by further integrating the spatial similarity-aided refinement, which completes the BFIAttack, the ASR reaches 73\%. These results indicate that all components of BFIAttack play a vital role in enhancing attack success rate.
\section{Countermeasures}


In this section, we discuss lessons learned and propose possible countermeasures to mitigate BFIAttack.





\textbf{Leveraging Multiple Wi-Fi Packets to Build Security Applications.}
Our evaluation reveals that BFIAttack exhibits a lower ASR for user authentication compared to other Wi-Fi-based physical-layer security applications. This is primarily because user authentication typically aggregates information from multiple Wi-Fi packets (e.g., 50 packets). Attempting to reconstruct a larger number of packets could decrease the likelihood of a successful attack. To validate this observation, we conduct experiments that extend the number of packets used in the attack. Specifically, applying BFIAttack over 200 packets results in a reduced ASR of 50.1\%, and further increasing the number of packets to 500 packets leads to a lower ASR of 35.0\%. These findings indicate that aggregating more packets enhances robustness against BFIAttack. However, this improvement in security comes at the cost of increased Wi-Fi-based system overhead, highlighting a tradeoff between robustness and efficiency in Wi-Fi-based system design.

\textbf{Incorporating CSI Phase Information.}
Incorporating CSI phase information into Wi-Fi-based security applications is a potential strategy to mitigate BFIAttack. This is motivated by two key observations. First, as indicated by Equation~\eqref{equ14}, the closed-form CSI reconstruction can only recover the relative phase between antenna elements, which may be insufficient for spoofing applications that leverage phase information. Second, according to Equation~\eqref{eq:opt_loss}, which contains $|T|$, the process inherently discards phase information, leading to potential discrepancies in the reconstructed CSI phase.
Despite its defensive potential, integrating the CSI phase into security applications requires precise phase calibration, which may introduce additional system complexity and computational overhead. 

\textbf{Avoid Single-Antenna Wi-Fi Devices.}
The most straightforward countermeasure is to avoid using Wi-Fi devices equipped with a single antenna, particularly those compliant with IEEE 802.11ac/ax standards. In such scenarios, CSI can be almost perfectly reconstructed from BFI, resulting in an extremely high attack success rate with only a single attack attempt.
In contrast, multi-antenna Wi-Fi devices pose greater challenges to attacks, which can be partially mitigated by leveraging more packets and phase information. From a practical deployment perspective, prioritizing multi-antenna Wi-Fi devices can therefore enhance resilience against BFIAttack.

\section{Discussion}

\textbf{Environmental Changes Affect Both CSI-Based Security Applications and BFIAttack.}
In our evaluations, we assume that the environment remains unchanged after the Wi-Fi/CSI-based security applications are established. If the environment changes, the multipath conditions also change, rendering the security applications invalid, which is an inherent challenge of Wi-Fi sensing~\cite{tan2022commodity}. Under the same conditions, BFIAttack likewise becomes ineffective. Nevertheless, a new BFIAttack can be launched once the user or device rebuilds the security applications in the new environment.

\textbf{Attack on Other Wi-Fi CSI-based Physical-Layer Security Applications.}
While this work focuses on three representative Wi-Fi CSI-based security applications, BFIAttack could be extended to compromise other CSI-based physical-layer security applications. One such example is Wi-Fi-based intrusion detection~\cite{lin2020revisiting}, where the presence of an intruder is inferred from perturbations in the wireless channel. In the future, we plan to adapt BFIAttack to counter such systems to render the intruder undetectable.


\textbf{Expanding the Attack Surface Enabled by BFI.}
BFI is a widely available and unencrypted form of channel information. It possesses significant adversarial potential that needs further exploration. BFI could be exploited to infer sensitive user information~\cite{xiao2025lend}, such as activity, location, and even fine-grained physiological signals (e.g., respiration and heartbeat). These capabilities introduce substantial privacy risks, particularly in IoT and smart environments. In future work, we will further expand and investigate the attack surface enabled by BFI.

\section{Conclusion}
In this work, we investigate the vulnerabilities of Wi-Fi-based physical-layer security by developing BFIAttack, a novel attack that reconstructs CSI from widely available plaintext BFI. BFIAttack effectively reconstructs the CSI of the legitimate user or device in both single-antenna and multi-antenna station scenarios using a closed-form solution and a maximum likelihood estimation-based method, respectively. We further introduce theory-practice dual constraints and spatial similarity-aided refinement to enhance attack performance in the multi-antenna scenario. Experimental results demonstrate that BFIAttack achieves high success rates and remains robust across diverse environments. Finally, we discuss potential countermeasures to mitigate the impact of BFIAttack.


\bibliographystyle{plain}
\bibliography{ref}

\end{document}